%% file: full2.tex
\documentclass[11pt,letterpaper]{article}
\usepackage[margin=1in]{geometry}

\usepackage{cite}
\usepackage{amsmath,amssymb,amsfonts,amsthm}
\usepackage{algorithmic}
\usepackage{graphicx,multicol}
\usepackage{textcomp}
\def\BibTeX{{\rm B\kern-.05em{\sc i\kern-.025em b}\kern-.08em
    T\kern-.1667em\lower.7ex\hbox{E}\kern-.125emX}}
\usepackage{epstopdf}
\usepackage{caption}
\usepackage{url}
\usepackage{cite}
\usepackage{color}
\usepackage{subcaption}
\usepackage{marvosym}

\usepackage{tikz}
\usetikzlibrary{shapes,arrows}

\usepackage[authoryear]{natbib}
\usepackage[pdfpagelabels,pdfpagemode=None,breaklinks=true]{hyperref} \hypersetup{
     colorlinks   = true,
     citecolor    = blue
}
\renewcommand{\cite}{\citep}

\newtheorem{theorem}{Theorem}
\newtheorem{corollary}{Corollary}
\newtheorem{lemma}{Lemma}
\newtheorem{remark}{Remark}
\newtheorem{definition}{Definition}
\newtheorem{example}{Example}
\newtheorem{proposition}{Proposition}
\newtheorem{assumption}{Assumption}

\allowdisplaybreaks

\newcommand{\ignore}[1]{}
\newcommand{\todo}[1]{#1}
\DeclareMathOperator*{\argmin}{argmin}
\DeclareMathOperator*{\argmax}{argmax}

\begin{document}

\title{Controlling Human Utilization of Failure-Prone Systems via Taxes\thanks{This research was supported in part by the National Science Foundation, under grant CNS-1718637. A preliminary version of this work appeared in the proceedings of the IEEE Conference on Decision and Control, 2016 \cite{hota2016controlling}.}}
\author{~Ashish~R.~Hota~and~Shreyas~Sundaram\thanks{Ashish R. Hota is with the Department of Electrical Engineering, IIT Kharagpur, India. Shreyas Sundaram is with the School of Electrical and Computer Engineering, Purdue University, USA. (e-mail: ahota@.ee.iitkgp.ac.in, sundara2@purdue.edu)}}%

\date{}
\maketitle

\begin{abstract}
We consider a game-theoretic model where individuals compete over a shared failure-prone system or resource. We investigate the effectiveness of a taxation mechanism in controlling the utilization of the resource at the Nash equilibrium when the decision-makers have behavioral risk preferences, captured by {\it prospect theory}. We first observe that heterogeneous prospect-theoretic risk preferences can lead to counter-intuitive outcomes. In particular, for resources that exhibit network effects, utilization can increase under taxation and there may not exist a tax rate that achieves the socially optimal level of utilization. We identify conditions under which utilization is monotone and continuous, and then characterize the range of utilizations that can be achieved by a suitable choice of tax rate. We further show that resource utilization is higher when players are charged differentiated tax rates compared to the case when all players are charged an identical tax rate, under suitable assumptions.
\end{abstract}

\section{Introduction}
\label{section:cdc_introduction}
\input{introduction.tex}

\input{preliminary.tex}
\input{taxation.tex}
\input{continuity.tex}

\section{Uniform versus Differentiated Tax Rates}
\label{section:cdc_reference}
\input{ReferenceHeterogeneity.tex}

\section{Discussions}
\label{section:discussion}

\subsection{Recent applications of Fragile CPR games}
We note that a detailed illustration of the proposed taxation scheme in an application is beyond the scope of this paper. Nevertheless, we highlight two recent applications that have been modeled in the framework of Fragile CPR games subsequent to the publication of our earlier works \cite{hota2014fragility,hota2016controlling}.

\begin{itemize}
\item The authors in \cite{vamvakas2019controlling,vamvakas2019dynamic} consider a 5G non-orthogonal multiple access (NOMA) wireless network where users split their transmission powers via the licensed band (with a flat fee and guaranteed quality of service) and an unlicensed band. The latter is a {\it congestible resource} that can be used without restrictions, but is constrained, and hence subject to overexploitation and collapse. The authors model the competition between users in this setting via a Fragile CPR game, and largely rely on the proof techniques developed in our prior work \cite{hota2014fragility} to show the existence, uniqueness and convergence results pertaining to the pure Nash equilibrium. Furthermore, \cite{vamvakas2019controlling} numerically illustrates the performance of a quadratic pricing scheme to control the utilization of the unlicensed band modeled as the fragile resource.
\item In \cite{gupta2019achieving}, the authors consider a setting where a group of players collaborate to serve a set of consumers, and to review their quality of service. Players choose their rate of service and rate of review subject to constraints. While the return from {\it service tasks} is proportional to the service rate chosen by the player, the rate of return from {\it review tasks} is an increasing function of the total review rate chosen by all players. Furthermore, if the aggregate review rate chosen by the players is high, it could lead to insufficient service rates and possible delay in reviewing. The authors model this interaction as a Fragile CPR game with review tasks modeled as a CPR with an {\it increasing rate of return}, characterize the pure Nash equilibrium, and empirically analyze the price of anarchy.
\end{itemize}

Our findings on incentivizing users in Fragile CPR games to control the utilization of the CPR are applicable in both the above settings.

\subsection{Perspectives on incentive design}
As discussed in the introduction, our focus has been to identify conditions under which there exists a tax rate that achieves a desired level of utilization of the fragile CPR under decentralized decision-making. An equally pertinent question is how to compute taxes to achieve this desired level of utilization in both cases when the parameters in the utility functions of the players (such as $\alpha_i$ or $\gamma_i$) are known and when these parameters unknown/uncertain. This is a fairly fundamental question for this class of games, and is beyond the scope of this paper. We add the following discussion in this regard and highlight some avenues for future research.

Our approach can be viewed in the framework of {\it closed-loop or inverse Stackelberg games} studied in \cite{ho1981information,ho1982control}. In this paradigm, the social planner is viewed as a leader who announces a mapping that maps the actions of the agents into an incentive signal or tax, and the players are viewed as followers who choose their actions to minimize their payment-dependent cost functions. Similarly, in this paper, we consider tax payments that are proportional to the investments of the players (i.e., the mapping mentioned above is linear), and the goal is to understand whether a given level of {\it aggregate} utilization can be achieved at the equilibrium for this class of taxation schemes. Nevertheless, with the exception of specific choices of utility functions, there is no general theory (that we know of) for computing those incentives both when the utility functions (or parameters therein) of the players are known and when they are unknown/uncertain \cite{ratliff2019perspective}.

To the best of our knowledge, the only exception is a recent working paper \cite{ratliff2018adaptive} which assumed that the utility functions of the players are given by a linear combination of a set of basis functions with coefficients being the preferences of the players that are not known to the social planner. In this case, they showed that the social planner can choose a set of incentives, observe the decisions made by the players, and adjust the incentive signals such that the actions of the agents at the equilibrium and the corresponding incentives coincide with the values desired by the social planner. Nevertheless, there are two important differences between the problem studied in this paper and the setting in \cite{ratliff2018adaptive}. In our setting, (i) the utilities of the players are not linear in the parameters, and (ii) the social planner is primarily interested in the aggregate level of utilization as opposed to the individual decisions of the players. We envision that an adaptive taxation approach inspired by \cite{ratliff2018adaptive} can be designed and shown to converge to a desired level of utilization (provided that there exists a tax rate that can achieve the desired level of utilization in the first place, the conditions for which are derived in this paper).

Another potential approach would be to use feedback to compute the tax rate as a function of the desired aggregate outcome and the observed aggregate outcome. With a suitable design of taxes, the ``closed-loop system" would be such that its equilibrium point would coincide with the Nash equilibrium with utilization coinciding with the desired utilization. A similar approach was studied in \cite{guan2019cumulative} in the context of dynamic pricing in transportation systems, albeit for the case when there is a single decision-maker, and in \cite{barrera2014dynamic} for a class of congestion games. Extending this approach to the class of games studied here is a promising future research direction.

\section{Conclusion}
\label{section:cdc_conclusion}
We investigated the effectiveness of a taxation mechanism in controlling the utilization of a failure-prone shared resource under prospect-theoretic risk preferences of users. We first showed the existence and uniqueness of PNE in Fragile CPR games under taxation. We then showed that for resources that exhibit network effects, heterogeneous prospect-theoretic utilities of the players can lead to {\it increase} in utilization and fragility with higher tax rates, and the utilization at the Nash equilibrium can be discontinuous in the tax rate. In contrast, for resources with a decreasing rate of return or congestion effects, utilization is always decreasing and continuous in the tax rate. Building upon these insights, we identified the range of utilization that can be achieved by a suitable choice of tax rate for both classes of resources. Finally, we showed that for homogeneous loss averse players, imposing differentiated tax rates results in higher utilization compared to the case where all players are charged an identical tax rate. Our results highlight the nuances of controlling human behavior under uncertainty, and provide compelling insights on how to identify and control their utilization of shared systems via economic incentives. 

\section*{Acknowledgment}
We thank Prof. Siddharth Garg (NYU) for helpful discussions, and the anonymous reviewers for their suggestions.

\bibliographystyle{plainnat}
\bibliography{refs}

\newpage

\appendix
\section*{Appendices}
\input{appendix_PNE.tex}
\input{appendix_network.tex}
\input{appendix_congestion.tex}
\input{appendix_diff.tex}

\end{document}

%% file: introduction.tex
Large-scale cyber-physical systems form the basis of much of society's critical infrastructure \cite{kim2012cyber}, and thus must be designed to be resilient to failures and attacks in order to avoid catastrophic social and economic consequences. While there are a variety of angles to designing such systems to be more resilient (including the design of secure control schemes \cite{teixeira2015secure,pasqualetti2013attack}, interconnection topologies \cite{buldyrev2010catastrophic,yagan2012optimal}, and resilient communication mechanisms \cite{jaggi2007resilient}), there is an increasing realization that the resilience of these systems also depends crucially on the {\it humans} that use them \cite{vanderhaegen2017towards,hota2017thesis}. Therefore, in order to design more resilient socio-cyber-physical systems, it is critical to understand (in a rigorous mathematical framework) the decisions made by humans in decentralized and uncertain environments, and to influence those decisions to obtain better outcomes for the entire system \cite{hota2017thesis,reverdy2014modeling,munir2013cyber}.

In this paper, we investigate the impacts of human decision-making on the resilience of a shared system in a game-theoretic framework. Game theory has emerged as a natural framework to investigate the impacts of decentralized decision-making on the efficiency, security and robustness of large-scale systems \cite{marden2015game,hota2017thesis}. When the utilities of the decision-makers or players are uncertain (e.g., due to risk of system failure or cyber-attack), their {\it risk preferences} play a significant role in shaping their behavior. With the exception of a few recent papers, most of the existing theoretical literature involving uncertainty models decision-makers as risk neutral (expectation maximizers) or risk averse (expected utility maximizers with respect to a concave utility function). However, empirical evidence has shown that the preferences of human decision-makers systematically deviate from the preferences of a risk neutral or risk averse decision-maker \cite{kahneman1979prospect,camerer2011advances}. Specifically, humans compare outcomes with a reference utility level, and exhibit different attitudes towards {\it gains} and {\it losses}. In their Nobel-prize winning work, Kahneman and Tversky proposed {\it prospect theory} \cite{kahneman1979prospect} in order to capture these attitudes with appropriately defined utility and probability weighting functions.\footnote{The probability weighting function captures the transformation of true probabilities into perceived probabilities by humans. We do not consider the impact of probability weighting in this work.} Prospect theory has been one of the most widely accepted models of human decision-making, and has shown its relevance in a broad range of disciplines \cite{camerer2011advances,barberis2013thirty,holmes2011management}, including recent applications in engineering \cite{hota2015interdependent,el2016prospect,nadendla2017strategic,etesami2018stochastic,nar:2017aa}. 

Motivated by the strong empirical and behavioral foundations of prospect theory, we study how to control the behavior of human decision-makers with prospect-theoretic utilities in a game-theoretic setting. We consider a broad class of games where users compete over a shared failure-prone system. We use the term ``resource" to refer to this shared system to maintain consistency with related game-theoretic models. Specifically, in our setting, a set of players split their budget between a safe resource with a constant return and a shared ``common pool" resource (CPR). As total investment or utilization by all players in the CPR increases, it becomes more likely for the CPR to fail, in which case the players do not receive any return from it. If the CPR does not fail, then the players receive a return per unit investment according to a rate of return function. Shared resources with increasing rates of return exhibit so-called {\it network effects} \cite{katz1994systems}; examples include online platforms for gaming, peer-to-peer file sharing systems, and social networks.\footnote{However, there are instances where authorities have shut down large online platforms that encourage illegal activities \cite{Wired2002,Johnson1459}. This is captured by resource failure in our setting.} CPRs with decreasing rates of return model {\it congestion effects} and describe engineered systems such as transportation and communication networks \cite{nisan2007algorithmic,orda1993competitive} and natural resources such as fisheries \cite{ostrom1994rules}. We consider CPRs with both network and congestion effects in this work. In Section \ref{section:fcprgame1}, we further discuss how this general model captures the externalities present in several applications. 

\subsection{Contributions}

We study a tax mechanism where each player is charged a tax amount proportional to her investment in the CPR. A central authority chooses the tax rate to control the utilization of the shared resource. Analysis of this taxation scheme is quite challenging under prospect-theoretic preferences. Building upon the analysis in \cite{hota2014fragility} (where we analyzed users' equilibrium strategies in the absence of taxation), we first show that the game admits a unique pure Nash equilibrium (PNE). We refer to the total investment in the CPR at a PNE as its {\it utilization}, and the failure probability as its {\it fragility}.

\todo{In particular, our focus on resource utilization is driven by the fact that it is an important metric relevant in many applications; e.g., the total traffic on a highway gives an indication of the level of congestion and throughput. Furthermore, in failure-prone systems, fragility also depends on the utilization rather than utility. In contrast with the total utility of all users, which is often used as a metric to capture the effects of decentralized decision-making \cite{nisan2007algorithmic}, utilization is agnostic to the behavioral risk preferences of the users.} Thus, our primary goal is to identify conditions under which:
\begin{enumerate}
\item there exists a tax rate that achieves a desired (e.g., socially optimal) level of CPR utilization, and
\item there exists an optimal tax rate that maximizes a continuous function of the tax rate and utilization (such as the revenue).
\end{enumerate} 

In order to answer these questions, we provide conditions under which utilization is monotone and continuous in the tax rate. It is perhaps natural to expect that a higher tax rate will reduce the utilization in a continuous manner. However, for CPRs that exhibit network effects, we find that behavioral risk preferences can sometimes cause utilization (and fragility) to {\it increase} with a higher tax rate. Furthermore, we illustrate that utilization can be discontinuous as the tax rate increases, both as a consequence of the shape of the utility function, and under heterogeneous prospect-theoretic preferences. We (separately) identify (i) conditions on the CPR characteristics and prospect-theoretic parameters under which utilization decreases monotonically with tax rate, and (ii) the range of tax rates over which the utilization varies continuously. 

In contrast to CPRs that exhibit network effects, we show that for CPRs that exhibit congestion effects, utilization is continuous and monotonically decreasing in the tax rate under general prospect-theoretic preferences of the players. Building upon these insights, we then identify the range of utilization that can be achieved via our taxation scheme. Finally we show that imposing different tax rates on a set of homogeneous loss averse players leads to a higher utilization than imposing a uniform tax rate (equal to the mean of the heterogeneous tax rates). \todo{In addition, when players have different sensitivities to taxes, imposing discriminatory taxes inversely proportional to the tax sensitivity parameters minimizes the utilization.}

\subsection{Related work}
Within the game-theoretic framework, controlling resource utilization levels through economic incentives such as taxes and rewards has been studied extensively \cite{delaney2015payments,brown2013social,swamy2012effectiveness}. In \cite{delaney2015payments}, the authors study how a taxation scheme known as Pigovian tax improves social welfare at a PNE in a CPR game. The effect of player-specific tax sensitivities on the price of anarchy were studied in \cite{brown2013social,brown2015robustness} in the context of nonatomic congestion games. In contrast, our game formulation is an instance of atomic splittable congestion games \cite{roughgarden2011local}. To the best of our knowledge, there has been no investigation of the impact of behavioral risk preferences on users' strategies under taxation in congestion or CPR games. 

%% file: preliminary.tex
\section{Prospect Theory}\label{section:PT}

As discussed in the previous section, our focus is on behavioral preferences captured by the utility function of {\it prospect theory}~\cite{kahneman1979prospect}. Specifically, consider a gamble that has an outcome with value $z \in \mathbb{R}$.  A prospect-theoretic individual {\it perceives} its utility in a skewed manner, via the function
\begin{equation}
u(z,z_0) =
\begin{cases}
(z-z_0)^\alpha, & \text{when }z \geq z_0 \\
-k(z_0-z)^\alpha & \text{otherwise},
\end{cases}
\label{eq:prospect}
\end{equation}
where $z_0$ is the reference point, $\alpha \in (0,1]$ is the {\it sensitivity parameter} and \todo{$k \in (0,\infty)$} is referred to as the {\it loss aversion index}. Increase in utility with respect to the reference point ($z \geq z_0$) is referred to as a {\it gain} and decrease in utility is referred to as a {\it loss} ($z < z_0$).

\begin{figure}[t]
	\centering
	\includegraphics[width=6cm,height=4cm]{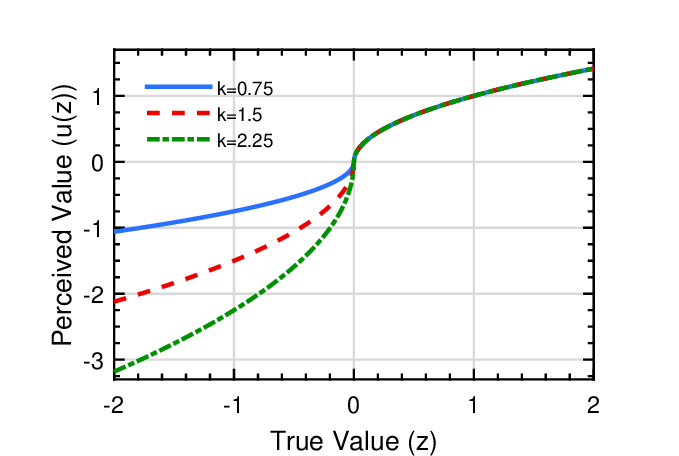}
	\caption{Prospect-theoretic utility function \eqref{eq:prospect} with $\alpha = 0.5$ and reference point $z_0 = 0$.}
	\label{fig:prospectvalue}
\end{figure}

The parameter $\alpha$ shapes the utility function according to observed behavior, i.e., the utility function is concave in the domain of gains and convex in the domain of losses. Accordingly, the decision maker is said to be ``risk averse" in gains and ``risk seeking" in losses. As its name indicates, the parameter $k$ captures loss aversion behavior. Specifically, when $\alpha=1$, a loss of \$$1$ {\it feels} like a loss of \$$k$ to the player. A value of $k>1$ implies that the individual is {\it loss averse}, while $k < 1$ implies that the individual is {\it gain seeking}. When the reference point is an exogenous constant, the values $k=1$ and $\alpha=1$ capture risk neutral behavior. A smaller $\alpha$ implies greater deviation from risk neutral behavior. The shape of the value function is shown for different values of $k$ in Figure~\ref{fig:prospectvalue}.

\section{Fragile Common Pool Resource Game}
\label{section:fcprgame1}

We start by introducing the Fragile Common Pool Resource game \cite{hota2014fragility}. Let $\mathcal{N}=\{1,2,\ldots,n\}$ be the set of players. Each player has an endowment or wealth equal to $1$ which she must split between a safe resource and a shared common pool resource (CPR). We define the strategy of a player $i \in \mathcal{N}$ as her investment in the CPR, denoted by $x_i \in [0,1]$. The total investment by all players in the CPR is denoted by $x_T = \sum_{i \in \mathcal{N}}x_i$. Following conventional notation, we denote the profile of investments by all players other than $i$ as ${\bf x}_{-i} \in [0,1]^{n-1}$. Furthermore, let $\bar{x}_{-i} = \sum_{j=1,j\neq i}^n x_j$, be the total investment of all players other than $i$.

Players receive returns on their investments from both resources. The return per unit investment from the safe resource is normalized to $1$, i.e., player $i$ investing $1-x_i$ in the safe resource receives a return of $1-x_i$. The return from the CPR is subject to risk, captured by a probability of failure $p(x_T)$, which is a function of the aggregate investment in the CPR. If the CPR fails, players do not receive any return from it. If the CPR does not fail, it has a per unit return that is a function of the total investment $x_T$, denoted by $\tilde{r}(x_T)$. In other words, player $i$ gets $x_i\tilde{r}(x_T)$ from the CPR when it does not fail.

The above formulation has been studied in many different contexts as described below.
\begin{enumerate}
\item The above formulation was studied as common pool resource games to model competition over failure-prone shared resources such as fisheries \cite{ostrom1994rules,walker1992probabilistic}.
\item CPR games, without resource failure, are equivalent to an instance of {\it atomic splittable congestion games} (studied in the context of traffic routing \cite{roughgarden2011local,orda1993competitive}) on a network with two nodes and two parallel links joining them. One link corresponds to the CPR described above and the second has a constant delay of $1$. 
\item Fragile CPR games are related to the setting in \cite{el2016prospect}, where players are microgrid operators who decide the fraction of energy to store for potentially selling at a higher price in the event of an emergency.\footnote{While the authors of \cite{el2016prospect} model microgrid operators as prospect-theoretic agents, the utilities are defined quite differently, and their objective is to study the effects of variations in reference points.} Both settings are related if we define the investment of a player as the fraction of stored energy, and $p(x_T)$ as the probability that the energy requirement during emergency is smaller than the total stored energy (i.e., energy price does not increase and the players incur losses).
\item In resource dilemma games \cite{budescu1995common}, players bid for utilizing a fraction of a shared resource with unknown size. If the total demand exceeds the size of the resource, no player receives any benefit. This model is potentially relevant when a set of users compete over a shared energy storage system \cite{paridari2015demand}. This class of games is closely related to Fragile CPR games where $x_i$ is the bid of player $i$, and $p(x_T)$ is the distribution of resource size. 
\end{enumerate}

In addition, two recent applications in the context of 5G wireless networks \cite{vamvakas2019controlling,vamvakas2019dynamic} and collaborative tasks \cite{gupta2019achieving} have been modeled in the framework of Fragile CPR games. We provide a brief discussion of those models in Section \ref{section:discussion}.

Given the breadth of applications where this formulation arises, the goal of this paper is to understand to what extent we can control the utilization ($x_T$) of the resource at the Nash equilibrium by imposing taxes on players' investments.

%% file: taxation.tex
\section{Prospect-Theoretic Utility and Pure Nash Equilibrium under Taxation}
\label{section:fcprgame}

\tikzstyle{block} = [draw, fill=red!20, rectangle, align=center, minimum width=2.5em,
    minimum height=3.5em]
\tikzstyle{block1} = [draw, fill=red!20, rectangle,  
    minimum height=1em, minimum width=1em]
\tikzstyle{sum} = [draw, fill=blue!20, circle, node distance=1cm]
\tikzstyle{input} = [coordinate]
\tikzstyle{output} = [coordinate]
\tikzstyle{pinstyle} = [pin edge={to-,thin,black}]

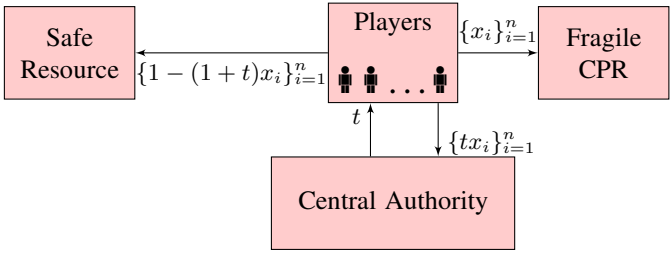
\begin{figure}[t]
\centering
\begin{tikzpicture}[auto, node distance=2cm,>=latex']
    \node [block] (controller) {Players \\ \\ \Large{\Gentsroom} \Large{\Gentsroom} \ldots \Large{\Gentsroom}};
    \node [block, right of=controller, text width=1.5cm,
            node distance=4cm] (system) {Fragile CPR};
    \node [block, left of=controller, text width=2cm, node distance=4.3cm] (safe) {Safe Resource};
    \node [block, below of=controller, text width=3cm, node distance=2.5cm] (measurements) {Central Authority};
    \draw [->] (controller) -- node[name=u1] {\small{$\{x_i\}^n_{i=1}$}} (system);
    \draw [->] (controller) -- node[name=u2] {\small{$\{1-x_i\}^n_{i=1}$}} (safe);
    \draw [->] ([xshift=-0.3cm]measurements.north) -- node [near end] {\small{$t$}} ([xshift=-0.3cm]controller.south);
    \draw [->] ([xshift=0.6cm]controller.south) -- node [near end] {\small{$\{tx_i\}^n_{i=1}$}} ([xshift=0.6cm]measurements.north);
\end{tikzpicture}
\caption{Central authority sets a tax rate $t$ to control human utilization of a failure-prone common pool resource (CPR).}
\label{fig:control}
\end{figure}

We first consider the case where a central authority imposes a uniform tax rate $t \geq 0$ per unit investment in the CPR on the players. Figure \ref{fig:control} represents the schematic of our setting.

Under this taxation scheme, a player $i$ with investment $x_i \in [0,1]$ in the CPR is charged $tx_i$ as tax. We will consider the implications of player-specific tax rates in Section \ref{section:cdc_reference}. Each player is prospect-theoretic, with a player-specific loss aversion index \todo{$k_i \in (0,\infty)$} and sensitivity parameter $\alpha_i \in (0,1]$. We define the reference utility of a player $i$ as her utility when she invests entirely in the safe resource, i.e., chooses $x_i=0$. Accordingly, the reference utility is $1$ for every player. Now consider a strategy profile $\{x_j\}_{j\in\mathcal{N}}$ with total investment $x_T$. In the event of CPR failure, each player $i$ with a nonzero $x_i$ experiences a loss $-(1+t)x_i$, which comprises of the lost income from not investing $x_i$ in the safe resource, and the tax payment. If the CPR succeeds, the reference-dependent return is $x_i(\tilde{r}(x_T)-1-t)$, which could be positive (representing a gain) or negative (representing a loss) depending on the values of $t$ and $x_T$. For ease of exposition, we define $r(x_T) := \tilde{r}(x_T)-1$, and henceforth refer to $r(x_T)$ as the {\it rate of return function}.

Using the prospect-theoretic utility function \eqref{eq:prospect}, player $i$'s {\it perception} of gains and losses is
\begin{equation}\footnotesize{
u_{i}(x_i,{\bf x}_{-i}) :=
\begin{cases}
x_{i}^{\alpha_i}[(\max(r(x_T)-t,0))^{\alpha_i} \\
\qquad -k_i(-\min(r(x_T)-t,0))^{\alpha_i}], \text{w.p. } 1-p(x_T), \\
-k_i(1+t)^{\alpha_i}x^{\alpha_i}_i, \text{w.p. } p(x_T).
\end{cases}
\label{eq:utilitydef}}
\end{equation}
Player $i$ maximizes the expected utility with respect to the above utility function given by
\begin{align}
\mathbb{E}(u_{i}(x_i,{\bf x}_{-i})) = x^{\alpha_i}_i f_i(x_T,t), \label{eq:taxedutility1}
\end{align}
where
\begin{equation}\footnotesize{
f_i(x_T,t) :=
\begin{cases}
(r(x_T)-t)^{\alpha_i}(1-p(x_T))-k_i(1+t)^{\alpha_i}p(x_T),
\\ \qquad \qquad \qquad \quad \text{when} \quad r(x_T)-t \geq 0,
\\ -k_i\left[(t-r(x_T))^{\alpha_i}(1-p(x_T))+(1+t)^{\alpha_i}p(x_T)\right],
\\ \qquad \qquad \qquad \quad \text{otherwise}.
\end{cases}}
\label{eq:taxedutility}
\end{equation}
We refer to $f_i(x_T,t)$ as the {\it effective rate of return} of player $i$. The shapes of $f_i(x_T,t)$ for different parameters are shown in Figure \ref{fig:effror} in Appendix \ref{appendix:PNE}. We denote this class of Fragile CPR games as $\Gamma(\mathcal{N},\{u_i\}_{i \in \mathcal{N}})$. In this paper, we consider Fragile CPR games under the following assumptions.

\begin{assumption} \label{assumption:CDCI}
The class of Fragile CPR games $\Gamma(\mathcal{N},\{u_i\}_{i \in \mathcal{N}})$ has the following properties.
\begin{enumerate}
\item The failure probability $p(x_T)$ is convex, strictly increasing and continuously differentiable for $x_T \in [0,1)$ and $p(x_T)=1$ for $x_T \geq 1$.
\item The rate of return $r(x_T)$ is concave, positive, strictly monotonic and continuously differentiable.
\item Define
\begin{equation} \bar{t} := \sup\{t \geq 0| \max_{i \in \mathcal{N}} \max_{x_T \in [0,1]} f_{i}(x_T,t) > 0\}. \label{eq:def_bart} \end{equation}
We assume that $\bar{t} > 0$, and the tax rate $t \in [0,\bar{t})$.
\end{enumerate}
\end{assumption}
These assumptions capture a fairly broad class of resources, while retaining analytical tractability.

To explain the third point in Assumption \ref{assumption:CDCI}, note from the definition of $\bar{t}$ that for any tax rate $t \geq \bar{t}$, the effective rate of return is nonpositive for every player and every $x_T \in [0,1]$. Accordingly, all players invest $0$ in the CPR at any PNE. On the other hand, for $t < \bar{t}$, there exist player(s) who make a nonzero investment leading to nontrivial PNE investments. 

\begin{remark}
The taxation scheme introduced here can be viewed as a subsidy on the safe resource (which increases the rate of return of the safe resource to $1+t$). The reference-dependent utility under this subsidy is identical to \eqref{eq:utilitydef}. Such a subsidy scheme was studied in \cite{delaney2015payments} outside of the context of behavioral decision-making.
\end{remark}

We now establish the existence and uniqueness of PNE in Fragile CPR games under taxes.

\begin{proposition}\label{theorem:cdc_PNEexistence}
Consider a Fragile CPR game with a fixed tax rate $t$ satisfying Assumption~\ref{assumption:CDCI}. Then there exists a unique joint strategy profile $\{x^*_i\}_{i \in \mathcal{N}}$ which is a PNE.
\end{proposition}

The proof is analogous to the PNE characterization established in \cite{hota2014fragility}. The details are presented in Appendix \ref{appendix:PNE}. The proof of existence of a PNE is based on Brouwer's fixed point theorem, while the uniqueness result follows from certain structural properties of the best response map. 

At a given tax rate $t$, we denote the total investment in the CPR at the corresponding PNE as $x^t_{\mathtt{NE}}$, and refer to it as the {\it utilization} (of the CPR). Furthermore, we refer to the corresponding failure probability $p(x^t_{\mathtt{NE}})$ as its {\it fragility}. With a slight abuse of notation, we sometimes denote $x^t_{\mathtt{NE}}$ as a function of $t$, i.e., we let $x_{\mathtt{NE}}: [0,\bar{t}) \to [0,1]$ denote a function such that $x_{\mathtt{NE}}(t) := x^t_{\mathtt{NE}}$.

\subsection{Social Welfare}

As discussed in the introduction, one of the key motivations behind this work is to identify conditions under which a socially desired level of utilization can be achieved under decentralized decision-making via taxation. In the game theory literature \cite{nisan2007algorithmic}, a metric that is often used to capture a socially desired level of utilization is the resource utilization that maximizes the sum of utilities of all players (also referred to as the social welfare). Formally, the social welfare at a joint strategy profile $\mathbf{x} \in [0,1]^{n}$ and a given tax rate $t \in [0,\bar{t})$ is defined as
\begin{equation}
\Psi(\mathbf{x},t) = \sum_{i \in \mathcal{N}} \mathbb{E}u_i(x_i,\mathbf{x}_{-i}) = \sum_{i \in \mathcal{N}} x^{\alpha_i}_if_i(x_T,t),
\end{equation}
where $u_i$ is defined in \eqref{eq:utilitydef}. Due to the continuity of $\Psi$, there always exists a social welfare maximizing set of investments. The following result shows that the CPR utilization and fragility are higher at the PNE compared to their counterparts at a social welfare maximizing strategy profile.

\begin{proposition}\label{proposition:socopt}
\todo{For $t \in [0,\bar{t})$, let $\mathbf{x}^t_{\mathtt{OPT}}$ be a joint investment profile that maximizes $\Psi(\mathbf{x},t)$. Then, the resulting total CPR investment at the social optimum $x^t_{\mathtt{OPT}}$ satisfies $x^t_{\mathtt{OPT}} \leq x^t_{\mathtt{NE}}$.}
\end{proposition}

The result holds under general (heterogeneous) prospect-theoretic preferences of the users. We refer to Appendix \ref{appendix:PNE} for the proof. In particular, we have $x^0_{\mathtt{OPT}} \leq x^0_{\mathtt{NE}}$.

\todo{In the context of congestion games and CPR games, the utilization that maximizes the social welfare in the absence of taxation, i.e., $x^0_{\mathtt{OPT}}$, is often treated as a socially desired level of utilization. Indeed, in the context of transportation networks, this quantity represents the traffic flow that minimizes the total congestion for all users. Existing literature, such as \cite{brown2013social,swamy2012effectiveness,fotakis2010existence}, has primarily investigated the existence and computation of taxes such that the utilization at the PNE under taxes equals $x^0_{\mathtt{OPT}}$ in the absence of resource failure and behavioral decision-making. In the following section, we investigate this in Fragile CPR games and under prospect-theoretic preferences.}

%% file: continuity.tex
\section{Main Results}

Recall from the introduction that our goal is to characterize the range of utilizations (including $x^0_{\mathtt{OPT}}$) that can be achieved at the PNE by a suitable choice of tax rate. We provide this characterization in this section.  

\subsection{CPRs with network effects}

\begin{figure*}[t!]
    \begin{subfigure}[t]{0.32\textwidth}
            \centering
    \includegraphics[width=\linewidth]{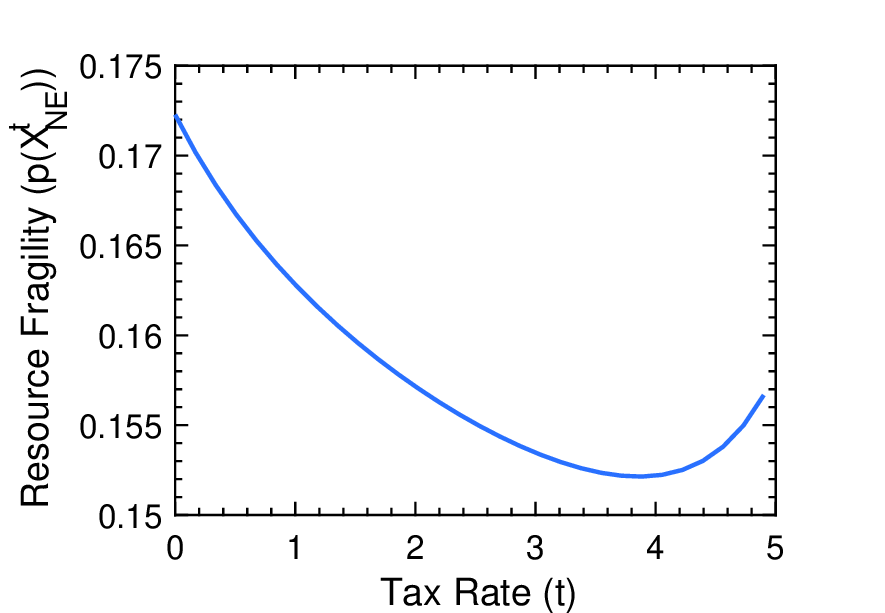}
    \caption{In Example \ref{example:alpha}, fragility is not monotone w.r.t. the tax rate ($k=1.2$ and $\alpha=0.15$)}
    \label{fig:taxationalpha}
    \end{subfigure}
    \hspace*{\fill}
    \begin{subfigure}[t]{0.32\textwidth}
        \centering
        \includegraphics[width=\linewidth]{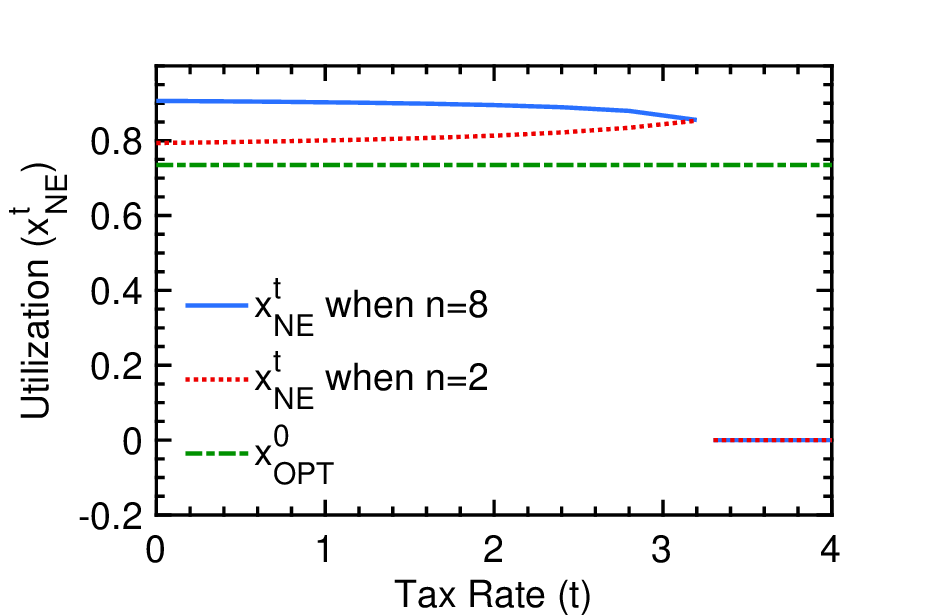}
        \caption{In Example \ref{ex:cont_seeking}, utilization is not continuous in the tax rate. There does not exist a tax rate which achieves the socially optimal level of utilization.}
        \label{fig:seeking_tax}
    \end{subfigure}
    \hspace*{\fill}
    \begin{subfigure}[t]{0.32\textwidth}
        \centering
        \includegraphics[width=\linewidth]{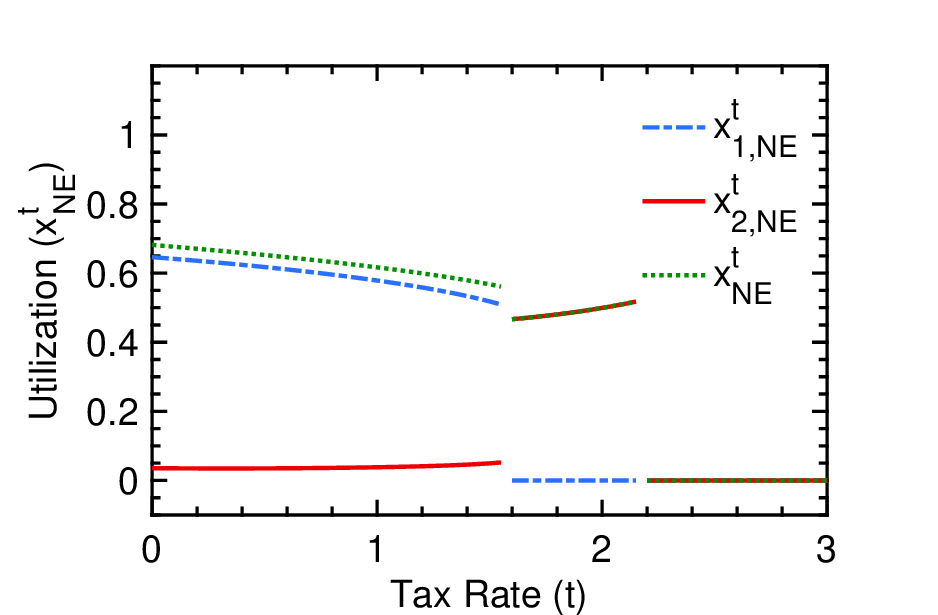}
        \caption{In Example \ref{ex:cont_multiple}, there are two points of discontinuity due to players with heterogeneous preferences.}
        \label{fig:multiple_tax}
    \end{subfigure}
    \caption{Illustration of lack of monotonicity and continuity of utilization and fragility in the tax rate under network effects.}
\end{figure*}

We first investigate CPRs with increasing rates of return. Recall that online platforms such as peer-to-peer file sharing systems are instances of CPRs that exhibit network effects. Note from \eqref{eq:taxedutility1} that the utility of a player in a game with a tax rate $t$ is equivalent to that in a game without taxes, but with a smaller rate of return function $(r(x_T)-t)$, and a larger index of loss aversion $k_i(1+t)^{\alpha_i}$. Therefore, intuition suggests that an increase in tax rate would lead to smaller utilization and fragility. However, the following example shows that imposing a higher tax rate can lead to higher utilization and fragility (at the PNE) under prospect theory and network effects.

\begin{example}\label{example:alpha}
Consider a Fragile CPR game with $n=3$ players. Let $r(x_T)=8x_T+5$ and $p(x_T)=x_T$. Let $\alpha=0.15$ and $k = 1.2$ for all players, i.e., all players are loss averse, and the deviation from risk neutral behavior ($\alpha=1$ and $k=1$) is significant. As shown in Figure~\ref{fig:taxationalpha}, when $t$ increases from $0$ to $4.9$, the fragility is not monotonically decreasing. Since $p(x_T)=x_T$, fragility equals utilization in this case.
\end{example}

\begin{remark}
In the above and subsequent examples, we compute the PNE strategy profile via sequential best response dynamics, which as argued in \cite{hota2014fragility}, converges to the PNE in this class of games. After convergence, we also verified that the strategy profiles satisfy the necessary optimality conditions.
\end{remark}

Recall from Figure~\ref{fig:prospectvalue} that $\alpha<1$ gives rise to risk seeking behavior in losses and risk averse behavior in gains. When the value of $\alpha$ is close to $0$, the modified loss aversion index $k(1+t)^\alpha$ does not increase by much at a higher tax rate. This encourages players to increase their investment into the CPR. Such behavior is not limited to the case when $\alpha$ is very small. In the conference version of this paper \cite{hota2016controlling}, we showed that a higher tax rate can lead to higher utilization when $\alpha=1$ and $k<1$. In both instances, players increase their investments to receive a higher return from the CPR and compensate for the tax payment (at the cost of increased risk of resource failure).

In addition to the general lack of monotonicity, we now illustrate that utilization can be discontinuous in the tax rate. 

\begin{example}\label{ex:cont_seeking}
Consider a Fragile CPR game with $p(x_T)=0.2+0.8x_T^4$, and $r(x_T) = 3x_T+1$. We consider homogeneous players with $\alpha=1, k=0.05$. As shown in Figure \ref{fig:seeking_tax} for $n=2$, $x^t_{\mathtt{NE}}$ increases continuously as $t$ increases until $\bar{t} = 3.21$, and then drops to $0$. On the other hand, when $n=8$, $x^t_{\mathtt{NE}}$ is decreasing in $t$, and once again, has a discontinuous jump at $\bar{t}$. The socially optimal utilization $x^0_{\mathtt{OPT}} = 0.7351$ is not achieved at any tax rate as shown in the figure.
\end{example}

In the above example, players were homogeneous with $k < 1$. Note that $x^t_{\mathtt{NE}}$ remained continuous for $t \in [0,\bar{t})$, but there did not exist a tax rate that could achieve the socially optimal level of utilization at a PNE. We now show that heterogeneity in prospect-theoretic preferences can induce discontinuity at tax rates smaller than $\bar{t}$.

\begin{example}\label{ex:cont_multiple}
Consider a Fragile CPR game with $p(x_T)=0.2+0.8x_T^4$ and $r(x_T) = 3x_T+1$, as before. Let there be two players, with prospect-theoretic parameters $\alpha_1=1, k_1=1.1$ and $\alpha_2 = 0.3, k_2 = 1.5$. In this case, $\bar{t} = 2.19$. Figure \ref{fig:multiple_tax} shows that for $t < 1.583$, player $1$ has a larger investment than player $2$. As $t$ becomes slightly larger than $1.583$, the investment by player $1$ drops to $0$, while player $2$ increases her investment. However, the total investment has a discontinuous jump from $0.5612$ to $0.4667$. As $t$ increases from $1.583$ to $2.19$, the investment by player $2$ increases continuously, and at $t = 2.19$, her investment drops to $0$.
\end{example}

Motivated by the above observations, we now identify conditions under which $x^t_{\mathtt{NE}}$ is monotone and continuous in $t$. We first introduce some notation. Let $a_t = 0$ if $r(0) \geq t$; otherwise, let $a_t \in [0,1)$ be the unique investment such that $r(a_t)=t$.\footnote{Since $t < \bar{t}$, there exists a player $i$ such that $\max_{x_T \in [0,1]} f_i(x_T,t) > 0$. Therefore, we must have $r(1) \geq t$.} For $t \in  [0,\bar{t}), x_T \in (a_t,1]$, let
\begin{align}\label{eq:cdc_function_inc_Q}
q_i(x_T,t) & := \frac{r'(x_T)(1-p(x_T))^2}{(r(x_T)+1)p'(x_T) - \alpha_i r'(x_T)(1-p(x_T))p(x_T)} \nonumber
\\ & \qquad \times \left(\frac{1+t}{r(x_T)-t}\right)^{1-\alpha_i}.
\end{align}
We show that if the index of loss aversion is larger than $q_i(x_T,t)$ for every player $i$ at suitable values of $x_T$ and $t$ identified below, then utilization is monotone in $t$.

In addition, let $\bar{t}_i := \sup\{t\geq0|\max_{x_T \in [a_t,1]} f_i(x_T,t)>0\}$, i.e., $\bar{t}_i$ is the highest tax rate such that player $i$ makes a nonzero investment in the CPR when investing in isolation. Recall from Appendix \ref{appendix:PNE} that $z^t_i := \argmax_{x_T \in [a_t,1]} f_i(x_T,t)$ is defined as the unique maximizer of $f_i(x_T,t)$ under network effects. We now state the following main result.

\begin{theorem}\label{theorem:cdc_inc_socopt}
Consider a Fragile CPR game satisfying Assumption \ref{assumption:CDCI} with an increasing $r(x_T)$ and player-specific prospect-theoretic preferences. 
\begin{enumerate}
\item Let $0 \leq t_2 < t_1 < \bar{t}$ with $x_{\mathtt{NE}}(t_2) > a_{t_1}$. Suppose $k_i > q_i(x_{\mathtt{NE}}(t_2),t_1) > 0$ for every player $i$.  Then, $x_{\mathtt{NE}}(t_1) \leq x_{\mathtt{NE}}(t_2)$.
\item $x_{\mathtt{NE}}(t)$ is continuous in $t$ for $t\in[0,\min_{i \in \mathcal{N}} \bar{t}_i)$.
\item For all continuous functions $w(x_{\mathtt{NE}}(t),t)$ and constants $\delta \in (0,\min_{i \in \mathcal{N}} \bar{t}_i)$, there exists a $t^* \in [0,\min_{i \in \mathcal{N}} \bar{t}_i-\delta]$ that maximizes $w(x_{\mathtt{NE}}(t),t)$ over $[0,\min_{i \in \mathcal{N}} \bar{t}_i-\delta]$.
\item Let $j \in \argmin_{i\in\mathcal{N}} \bar{t}_i$, and let $\bar{x}_j := \lim_{t \uparrow \bar{t}_j} z^t_j$. If $x_{\mathtt{NE}}(0) > \bar{x}_j$ (respectively, $x_{\mathtt{NE}}(0) < \bar{x}_j$), then for any given level of utilization $x^*\in(\bar{x}_j,x_{\mathtt{NE}}(0)]$ (respectively, $x^*\in[x_{\mathtt{NE}}(0),\bar{x}_j)$) there exists a tax rate $t$ such that $x^* = x_{\mathtt{NE}}(t)$.
\item If $\bar{x}_j < x^0_{\mathtt{OPT}}$, then there exists a tax rate $t^*$ such that $x_{\mathtt{NE}}(t^*)= x^0_{\mathtt{OPT}}$.
\end{enumerate}
\end{theorem}

The proof is presented in Appendix \ref{appendix:network}. We now describe several implications of the above result. The first statement is a condition that we can check to ensure that a higher tax rate will lead to smaller utilization. Furthermore, when all players have identical $\alpha$, we only need to check the condition for the player with the smallest loss aversion index.\footnote{In \cite{hota2014fragility}, we showed that when players have identical $\alpha$, the player with the smallest loss aversion index always has the largest investment at the PNE.}

The second statement guarantees that utilization remains continuous over a subset of tax rates for which utilization is nonzero. When all players have identical $\alpha$, the player with the largest loss aversion index has the smallest $\bar{t}_i$. When all players have identical $\alpha$ and $k$, $\bar{t}_i$ is identical for every player, and thus $\bar{t} = \min_{i \in \mathcal{N}} \bar{t}_i$. Thus, the conclusions of the above result holds over the entire range of tax rates over which PNE utilization is nonzero as is the case in Example \ref{ex:cont_seeking}. In contrast, in Example \ref{ex:cont_multiple}, the players had heterogeneous preferences with $\bar{t}_1 = 1.583$, while $\bar{t}_2 = 2.19$. As shown in Figure \ref{fig:multiple_tax}, utilization is continuous for $t \in [0,1.583)$ in accordance with the above result, and has a discontinuous jump at $t=1.583$. 

Finally, let all players have homogeneous preferences with $\alpha=1$ and $k \geq 1$. Recall that in this case, the prospect-theoretic utility \eqref{eq:prospect} is either linear or piecewise concave, and reflects risk neutral or risk averse preferences. We have the following corollary whose proof is stated in Appendix \ref{appendix:network}.

\begin{corollary}\label{corollary:cdc_inc_socopt_classical}
Let all players have $\alpha=1$ and $k \in [1,\infty)$. Let $0 \leq t_2 < t_1 < \bar{t}$. Then, $x_{\mathtt{NE}}(t_1) \leq x_{\mathtt{NE}}(t_2)$. Furthermore, there exists a tax rate $t^*$ such that $x_ {\mathtt{NE}}(t^*) = x^0_{\mathtt{OPT}}$.
\end{corollary}

Thus, it follows that the lack of monotonicity observed earlier is a consequence of prospect-theoretic risk preferences.

\subsection{CPRs with congestion effects}

The counterpart of Theorem \ref{theorem:cdc_inc_socopt} is stronger for CPRs with congestion effects. In contrast with the observations in the above subsection, for resources with a decreasing $r(x_T)$, we show here that an increase in tax rate always leads to smaller utilization of the CPR. Furthermore, the total investment at the PNE is continuous in $t$ for $t \in [0,\bar{t}]$, i.e., the entire range of tax rates with nonzero utilization. The results hold when $k_i \in (0,\infty)$ and $\alpha_i \in (0,1]$ are player-specific. 

\begin{theorem}\label{theorem:main_dec}
Consider a Fragile CPR game satisfying Assumption \ref{assumption:CDCI} with a decreasing $r(x_T)$ and player-specific prospect-theoretic preferences. 
\begin{enumerate}
\item Let $0 \leq t_2 < t_1 < \bar{t}$. Then, $x_{\mathtt{NE}}(t_1) \leq x_{\mathtt{NE}}(t_2)$.
\item The function $x_{\mathtt{NE}}(t)$ is continuous in $t$ for $t\in[0,\bar{t}]$.
\item For all continuous functions $w(x_{\mathtt{NE}}(t),t)$, there exists a tax rate $t^* \in [0,\bar{t}]$ that maximizes $w(x_{\mathtt{NE}}(t),t)$ over $[0,\bar{t}]$.
\item For any given level of utilization $x^* \in [0,x^0_{\mathtt{NE}}]$, there exists a tax rate $t \in [0,\bar{t}]$ such that $x^* = x_{\mathtt{NE}}(t)$. Specifically, there exists a tax rate $t^*$ such that $x^0_{\mathtt{OPT}} = x_{\mathtt{NE}}(t^*)$. In addition, for any $x^* > x^0_{\mathtt{NE}}$, there does not exist a positive tax rate that achieves it.
\end{enumerate}
\end{theorem}

In other words, any desired utilization $x^* \in [0,x^0_{\mathtt{NE}}]$ can be achieved by an appropriate choice of tax rate. We present the formal proof in Appendix \ref{appendix:congestion}.

Our discussion thus far assumes that the central authority imposes an identical tax rate $t$ on every player. In the following section, we compare the utilization when the central authority imposes different tax rates on different players with the utilization under a uniform tax rate for all players.

%% file: ReferenceHeterogeneity.tex
In order to isolate the effects of differentiated tax rates, we assume that all players have identical loss aversion indices $k \in (1,\infty)$ and $\alpha = 1$. Let $\gamma_i \in [0,1]$ be the {\it tax sensitivity} of player $i$, and let $\hat{t}_i \geq 0$ be the tax rate imposed on player $i$ by the central authority. The tax sensitivity is an inherent property of the players: a player $i$ with sensitivity $\gamma_i$ perceives the tax rate $\hat{t}_i$ imposed on her as $\gamma_i \hat{t}_i$. We define $t_i := \gamma_i \hat{t}_i$ as the {\it effective tax rate} experienced by player $i$. The expected utility only depends on the effective tax rate $t_i$ as shown below.

Let $S_{t_i} \subseteq [0,1]$ be the interval such that $r(x_T) - t_i \geq 0$ for $x_T \in S_{t_i}$. \todo{Following equation~\eqref{eq:taxedutility1}, the expected utility of player $i$ at a strategy profile with $x_T \in S_{t_i}$ is
\begin{align}
\mathbb{E}(u_i(x_i,\mathbf{x}_{-i})) & = x_i(r(x_T)-\gamma_i \hat{t}_i)(1-p(x_T)) \nonumber
\\ & \quad - k(1+\gamma_i \hat{t}_i) x_i p(x_T) \nonumber
\\ & = x_i f_i(x_T,t) =: x_i [\hat{f}(x_T) - t_i v(x_T)], \label{eq:taxsensitivity}
\end{align}
where $\hat{f}(x_T) := r(x_T)(1-p(x_T))-kp(x_T)$, and $v(x_T) := 1+(k-1)p(x_T)$ for $x_T \in [0,1]$.}

Impacts of tax sensitivities on price of anarchy in congestion games were studied recently in \cite{brown2013social}, outside of the context of behavioral risk attitudes. Under prospect theory, player-specific tax sensitivities can arise when players have different reference utilities. In particular, the utility in \eqref{eq:taxsensitivity} arises if the reference utility of player $i$ is $1-(1-\gamma_i) \hat{t}_i x_i$. In this case, player $i$ perceives her tax payment as part of her reference utility as opposed to treating it entirely as a loss. If $\gamma_i = 0$, the tax payment is included in the reference utility, and consequently, the results are same as the case without taxation.

\begin{remark}
Our results on PNE existence and uniqueness rely on the uniqueness, continuity and monotonicity properties of the best response. These properties remain unchanged with a linear scaling of the tax rate, and accordingly a PNE exists and is unique when the utilities are defined as in \eqref{eq:taxsensitivity}.
\end{remark}

Before we compare the PNE utilization of the CPR under uniform and player-specific tax rates, we first identify conditions for the existence of a uniform tax rate such that the PNE utilization under taxation is equal to the utilization at the social welfare maximizing solution under player-specific tax sensitivities. The following result is analogous to the prior results (Proposition \ref{proposition:socopt}, Theorem \ref{theorem:cdc_inc_socopt} and Theorem \ref{theorem:main_dec}) which did not consider tax sensitivities.

\begin{proposition}\label{prop:sensitivity_uniform}
Consider a Fragile CPR game satisfying Assumption \ref{assumption:CDCI} with $\alpha=1$ and $k>1$ for all players, player-specific tax sensitivity $\gamma_i \in [0,1]$, and a uniform tax rate $t \in [0,\bar{t})$ where $\bar{t} = \sup\{t\geq0|\max_{i \in \mathcal{N}} \max_{x \in [0,1]} f_i(x,t) > 0\}$ with $f_i$ defined in \eqref{eq:taxsensitivity}. Then, 
\begin{enumerate}
\item at a given $t \in [0,\bar{t})$, we have $x_{\mathtt{OPT}}(t) \leq x_{\mathtt{NE}}(t)$, and
\item for $0 \leq t_2 < t_1 < \bar{t}$, $x_{\mathtt{NE}}(t_1) < x_{\mathtt{NE}}(t_2)$.
\end{enumerate}
Furthermore, 
\begin{enumerate}
\item when $r(x_T)$ is increasing, $x_{\mathtt{NE}}(t)$ is continuous in $t$ for $t \in [0,t^*)$ where $t^* := \min_{i \in \mathcal{N}} \bar{t}_i, \bar{t}_i = \sup\{t\geq0|\max_{x \in [0,1]} f_i(x,t) > 0\}$, and if $x_\mathtt{OPT}(0) > x_\mathtt{NE}({t^*})$ then there exists a tax rate such that utilization at the NE is equal to $x_\mathtt{OPT}(0)$, and
\item when $r(x_T)$ is decreasing, $x_{\mathtt{NE}}(t)$ is continuous in $t$ for $t \in [0,\bar{t})$, and for any $x^* \in [0, x_{\mathtt{NE}}(0)]$, there exists a tax rate $t^*$ such that $x^* = x_{\mathtt{NE}}(t^*)$. 
\end{enumerate}
\end{proposition}

The proof is analogous to the proofs of Theorems \ref{theorem:cdc_inc_socopt} and \ref{theorem:main_dec}, and a sketch of the proof is presented in Appendix \ref{appendix:diff}.

We now establish that the PNE utilization is smaller when the effective tax rate is identical for all players compared to the utilization when the effective tax rates are heterogeneous, provided that the mean effective tax rates are identical. Consider the family $\mathbf{\Gamma}_m$ of Fragile CPR games with $n$ players each with $\alpha=1$ and $k>1$, $r(x_T)$ and $p(x_T)$ satisfying Assumption~\ref{assumption:CDCI}, and the mean of the effective tax rates being $t_m$. With a slight abuse of notation, we sometimes refer a player with effective tax rate $t_m$ as player $m$. Let $\bar{t}_m := \sup\{t\geq0|\max_{x_T \in S_{t_m}} f_m(x_T,t)>0\}$. The following result holds for both increasing and decreasing rate of return functions. 

\begin{proposition}\label{prop:diff}
Let $t_m \in [0,\bar{t}_m)$. Let $\Gamma_M \in \mathbf{\Gamma}_m$ be the game where the effective tax rate is $t_m$ for every player. Then, among all games in $\mathbf{\Gamma}_m$, CPR utilization is smallest in $\Gamma_M$.
\end{proposition}

The main ideas behind the proof are analogous to the ideas used in the proof of Theorem 5 in \cite{hota2014fragility}. The proof is presented in Appendix \ref{appendix:diff}. \todo{When the sensitivity parameters are known to the central authority, then the following corollary states the differentiated tax rates to be imposed on the players (subject to a mean tax rate constraint) in order to minimize the utilization.}

\begin{corollary}\label{corollary:diff}
\todo{Let $\Gamma$ be a Fragile CPR game with $n$ players each with $\alpha=1$, $k>1$, and satisfying Assumption \ref{assumption:CDCI}. Let the tax sensitivity of player $i$ be denoted as $\gamma_i$ with $\gamma_ i > 0$. Then, among all differentiated tax rates with a given mean tax rate $\hat{t}_m$, the choice $\hat{t}_i := \frac{n\hat{t}_m}{\gamma_i} \big(\sum^n_{i=1} \frac{1}{\gamma_i}\big)^{-1}, i \in \mathcal{N}$ minimizes the utilization at the PNE.}
\end{corollary}

The above result shows that when players are {\it loss averse} (i.e., $k > 1$) with $\alpha=1$, then the central authority should impose differentiated taxes {\it inversely proportional to their tax sensitivities} in order to minimize the utilization of the shared resource at the PNE. The proof is a straightforward consequence of Proposition \ref{prop:diff}. In particular, when players have identical sensitivity parameters, charging different tax rates to different players leads to higher utilization and fragility of the CPR. The analysis is significantly more involved when $\alpha < 1$. On the other hand, since the utilities are continuous in $\alpha$, we expect the above results to hold when $\alpha$ is close to $1$.

The results thus far assume that the players are homogeneous vis-a-vis their loss aversion indices. Preliminary investigations show that imposing a higher tax rate on users with smaller (or larger) loss aversion indices does not always lead to a smaller level of resource utilization. Thus, a counterpart of Corollary \ref{corollary:diff} does not hold when we consider heterogeneity in loss aversion instead of tax sensitivity. Further investigations on computing differentiated tax rates to minimize utilization (or any other objective such as maximizing revenue) remain open for future work (see the discussion below).

%% file: appendix_PNE.tex
\section{Characterization of Pure Nash Equilibrium and Social Optimum}\label{appendix:PNE}

In this section, we first prove the existence and uniqueness of a PNE in Fragile CPR games under taxation. We further show that the total investment in the CPR at a social optimum is at most that at the PNE. Specifically, we introduce several useful notations and preliminary results that are essential for subsequent analysis.

\subsection{Existence and Uniqueness of PNE}

We first describe the approach behind our analysis. \todo{We define the {\it best response} correspondence of a player $i$ as $B_i({\bf x}_{-i}) := \argmax_{x_i\in[0,1]} \mathbb{E}u_i(x_i,{\bf x}_{-i})$, where $\mathbb{E}u_i(\cdot)$ is defined in \eqref{eq:taxedutility1}. Let $B({\bf x}) := [B_1({\bf x}_{-1}), B_2({\bf x}_{-2}), \ldots, B_n({\bf x}_{-n})]$. We rely on the characterization that a joint strategy profile $\mathbf{x}^* = \{x_i^*\}_{i \in \mathcal{N}}$ is a PNE if and only if it is a fixed point of the best response map, i.e., ${\bf x^*} \in B({\bf x^*})$ \cite{ok2007real}. We show that a PNE exists by establishing the existence of a fixed point by applying Brouwer's fixed point theorem.} For this purpose, it is sufficient to show that $B_i$ is single-valued and continuous in ${\bf x}_{-i}$. The subsequent analysis follows in this direction.

We first introduce some relevant notation. Consider a Fragile CPR game with a fixed tax rate $t \in [0,\bar{t})$. Then, the PNE (if one exists) has nonzero CPR investments, and the total investment must be such that $r(x_T)-t \geq 0$ (from \eqref{eq:taxedutility}, we have $f_i(x_T,t) \geq 0 \implies r(x_T)-t \geq 0$). Accordingly, most of our analysis will focus on the range of total investments that lie within a subset $S_t \subseteq [0,1]$ such that $r(x_T)-t \geq 0$ for $x_T \in S_t$. When $r(x_T)$ is strictly decreasing, we have $S_t := [0,b_t]$, where
\begin{equation}
b_t :=
\begin{cases}
1, & \text{if } r(1)\geq t, \\
r^{-1}(t), & \text{if } r(1) < t, \\
\end{cases}
\label{eq:def_bt}
\end{equation}
where $r^{-1}(t)=\{y\in[0,1]|r(y)=t\}$. On the other hand, when $r(x_T)$ is strictly increasing, we have $S_t := [a_t,1]$, where
\begin{equation}
a_t :=
\begin{cases}
0, & \text{if } r(0)\geq t, \\
r^{-1}(t), & \text{if } r(0) < t. \\
\end{cases}
\label{eq:def_at}
\end{equation}
Note that for $t \in [0,\bar{t})$, $S_t$ is well defined and is nonempty.

We start with the following lemma. While the proof largely follows from identical arguments as the proof of Lemma 1 in \cite{hota2014fragility} (where we considered Fragile CPR games without taxation), we present it here as the proof formally defines several important quantities that are useful in the analysis throughout the paper. Recall that $\bar{x}_{-i}$ denotes the total investment by all players other than $i$.

\begin{figure*}
        \begin{subfigure}[t]{0.32\textwidth}
        \centering
        \includegraphics[width=\linewidth]{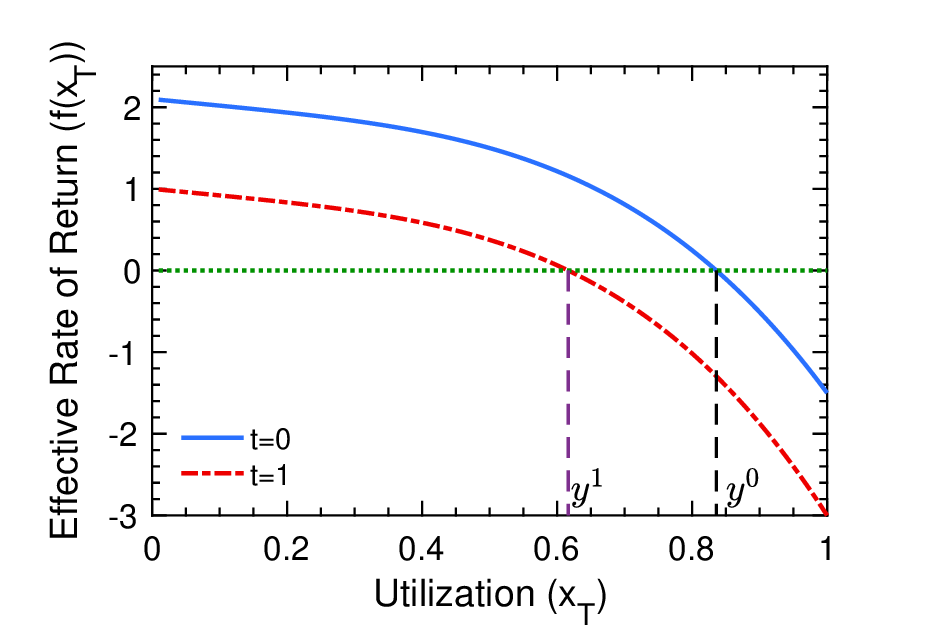}
        \caption{Effective rate of return under $r(x_T)=3-x_T$, $p(x_T)=0.2+0.8x_T^4$, $\alpha=1$ and $k=1.5$. Here $y^0 = 0.8359$ and $y^1 = 0.6166$.}
        \label{fig:neutral_eff}
    \end{subfigure}
    \hspace*{\fill}
    \begin{subfigure}[t]{0.32\textwidth}
        \centering
        \includegraphics[width=\linewidth]{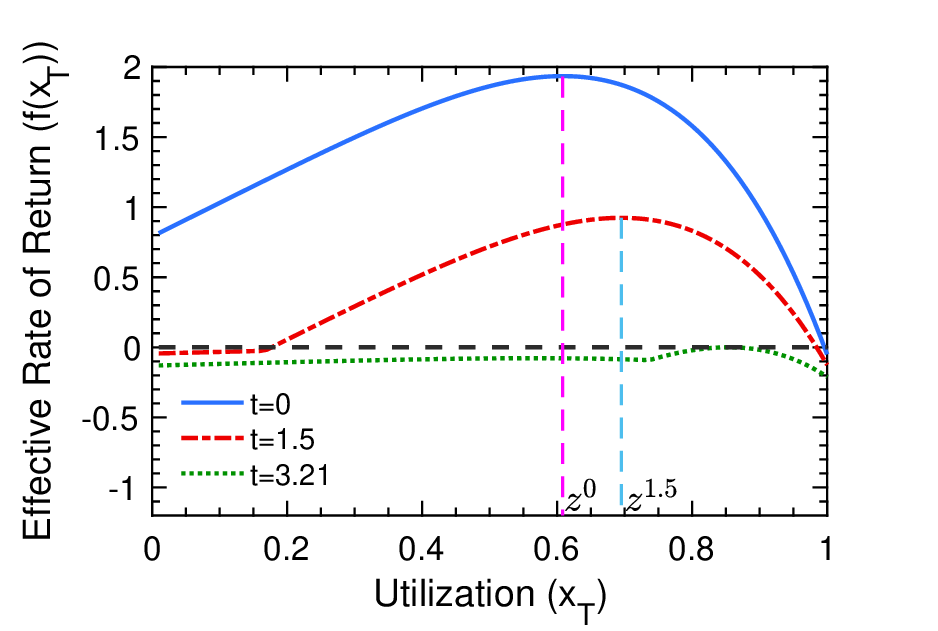}
        \caption{Effective rate of return under $r(x_T)=3x_T+1$, $p(x_T)=0.2+0.8x_T^4$, $\alpha=1$ and $k=0.05$. Here $y^0 = 0.9961$, $z^0=0.6083$, $y^{1.5}=0.9845$, $z^{1.5}=0.6952$, $y^{3.21}=0$, and $z^{3.21}=0.85$.}
        \label{fig:seeking_eff}
    \end{subfigure}
    \hspace*{\fill}
    \begin{subfigure}[t]{0.32\textwidth}
        \centering
        \includegraphics[width=\linewidth]{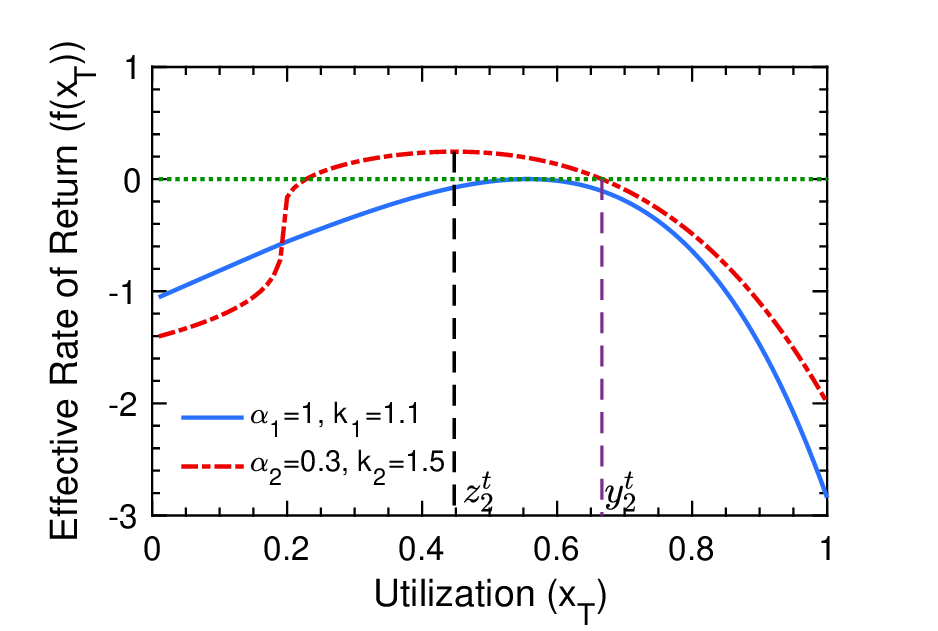}
        \caption{Effective rate of return for two players at $t=1.583$ under $r(x_T)=3x_T+1$, and $p(x_T)=0.2+0.8x_T^4$. Here $f_1(x_T,t) \leq 0$ for all $x_T \in [0,1]$, and $y^t_1=0$. Furthermore, $y^t_2 = 0.6663$ and $z^t_2 = 0.4476$.}
        \label{fig:multiple_eff}
    \end{subfigure}
    \caption{Shapes of the effective rate of return function under different CPR characteristics, risk preferences, and tax rates. The second argument $t$ is suppressed.}
    \label{fig:effror}
\end{figure*}

\begin{lemma}\label{lemma:interval} Consider a Fragile CPR game with a fixed tax rate $t \in [0,\bar{t})$. Then, for any player $i$, the following are true.
\begin{enumerate}
\item There exists a unique $y^t_{i} \in [0,1]$ such that if $\bar{x}_{-i} \geq y^t_{i}$, then $B(\mathbf{x}_{-i}) = \{0\}$. Furthermore, if $0 \in B(\mathbf{x}_{-i})$, then $\bar{x}_{-i} \geq y^t_{i}$. 
\item When $y^t_{i}>0$, $f_i(y^t_{i},t)=0$, and there exists an interval $\mathcal{I}^t_i \subseteq [0,y^t_{i}) \subset S_t$ such that if $\bar{x}_{-i} < y^t_{i}$, then each best response $b_i \in B_i({\bf x}_{-i})$ (i) is positive, and (ii)  satisfies $b_i + \bar{x}_{-i} \in \mathcal{I}^t_i$.
\item For $x_T \in \mathcal{I}^t_i$, we have $f_i(x_T,t)>0$ and $f_{i,x}(x_T,t) := \frac{\partial f_i(x_T,t)}{\partial x_T} <0$.
\end{enumerate}
\end{lemma}
\begin{proof} 
\todo{We first prove all three statements for CPRs with decreasing $r(x_T)$ in Case 1, and then consider CPRs with increasing $r(x_T)$ in Case 2.}

\noindent \textbf{Case 1:} {\bf $\mathbf{r(x_T)}$ is decreasing.} From the definition of $b_t$ in \eqref{eq:def_bt}, we obtain $f_i(b_t,t) < 0$ in \eqref{eq:taxedutility}. Straightforward calculation shows that $f_i(x_T,t)$ is strictly decreasing in $x_T$ when $x_T\in[0,b_t]$. If $f_i(0,t) \leq 0$, we define $y^t_i = 0$.

On the other hand, if $f_i(0,t) > 0$, we define $y^t_i \in S_t$ as the unique investment where $f_i(y^t_i,t) = 0$. If $\bar{x}_{-i} \geq y^t_{i}$, an investment $\epsilon > 0$ by player $i$ will lead to $f_i(\epsilon+\bar{x}_{-i},t) < 0$, and consequently a negative utility. Therefore, $B_i({\bf x}_{-i}) = \{0\}$ in this case. On the other hand, if $\bar{x}_{-i} < y^t_{i}$, there exists $\delta > 0$ such that $\delta+\bar{x}_{-i} < y^t_{i}$, and therefore, $f_i(\delta+\bar{x}_{-i},t) > 0$. Thus, the optimal CPR investment $x^*_i$ is nonzero and $x^*_i + \bar{x}_{-i} < y^t_i$. Accordingly, we define $\mathcal{I}^t_i := [0,y^t_{i})$. \todo{Since $f_i(x_T,t)$ is strictly decreasing in $x_T$ in this case, we have $f_i(x_T,t)>0$ and $f_{i,x}(x_T,t) <0$ for $x_T \in \mathcal{I}^t_i$.} 

\noindent \textbf{Case 2:} {\bf $\mathbf{r(x_T)}$ is increasing.} If $f_i(x_T,t) \leq 0$ for $x_T \in [a_t,1]$, we define $y^t_i = 0$, and $B_i({\bf x}_{-i}) := \{0\}$ for every ${\bf x}_{-i}$.

Now suppose there exists $x_T \in [a_t,1]$  where $f_i(x_T,t) > 0$. Straightforward calculation shows that $f_i(x_T,t)$ is strictly concave in $x_T$ when $x_T \in [a_t,1]$. Therefore, there exists a unique maximizer of $f_i(x_T,t)$ given by $z^t_i := \argmax_{x_T \in [a_t,1]} f_i(x_T,t)$. Note that we must have $z^t_i < 1$ since $f_i(1,t) < 0$. From the strict concavity of $f_i$, we have $f_{i,x}(x_T,t) := \frac{\partial f_i(x_T,t)}{\partial x_T} < 0$ for $x_T > z^t_i$. Thus, there exists a unique investment $y^t_i \in (z^t_i,1)$ such that $f_{i}(y^t_i,t) = 0$. \todo{In this case, we define $\mathcal{I}^t_i := (z^t_i,y^t_{i})$. Since $f_i(x_T,t)$ is strictly concave in $x_T$, and $z^t_i$ is its unique maximizer, we have $f_i(x_T,t) > 0$, and $f_{i,x}(x_T,t) <0$  for $x_T \in \mathcal{I}^t_i$.}

Now suppose the total investment by players other than $i$ satisfies $\bar{x}_{-i} \geq y^t_{i}$. Then any $x_i > 0$ would imply $f_i(x_i+\bar{x}_{-i})<0$, and $0$ is the unique best response. On the other hand, if $\bar{x}_{-i} < y^t_{i}$, there exists $\delta > 0$ such that $f_i(\delta+\bar{x}_{-i})>0$, and thus, all best responses must be positive. Note that we must necessarily have $\delta+\bar{x}_{-i} > a_t$. Now suppose $x^*_i \in B_i({\bf x}_{-i})$. Then it must necessarily satisfy the first order condition of optimality $\frac{\partial \mathbb{E}(u_i)}{\partial x_i} = 0$ for the utility in \eqref{eq:taxedutility1}, leading to
\begin{align}
& x^*_if_{i,x}(x^*_i+\bar{x}_{-i},t) + \alpha_i f_i(x^*_i+\bar{x}_{-i},t) = 0. \label{eq:def_foc}
\end{align}
Since $f_i(x^*_i+\bar{x}_{-i},t) > 0$, we must have $f_{i,x}(x^*_i+\bar{x}_{-i},t) < 0$, and therefore, $x^*_i+\bar{x}_{-i} \in \mathcal{I}^t_i$.
\end{proof}

\begin{remark}
Figure \ref{fig:effror} illustrates the quantities introduced in the above lemma; the subscript $i$ is dropped for convenience. Figure \ref{fig:neutral_eff} shows that $y^0 = 0.8359$ and $y^1 = 0.6166$ for a CPR with a decreasing rate of return. Note from the figure that $f(y^t,t)=0$ in both cases. Figure \ref{fig:seeking_eff} and \ref{fig:multiple_eff} show the values of $y^t_i$ and $z^t_i$ for a CPR with $r(x_T)=3x_T+1$ and $p(x_T)=0.2+0.8x^4_T$ for different tax rates and risk preferences. Note from the figures that $z^t_i$ is the maximizer of $f_i(x_T,t)$, and $f_i(y_i^t,t)=0$. The kinks in the last two figures occur at the respective $a_t$ values.
\end{remark}

We now build upon the above discussion, and introduce a few other important quantities. For a player $i$, we define
\begin{equation}\label{eq:defg}
g_i(x_T,t) := \frac{\alpha_i f_i(x_T,t)}{-f_{i,x}(x_T,t)}, \quad x_T \in S_t.
\end{equation}
It follows from the first order optimality condition in \eqref{eq:def_foc} that a nonzero best response $x^*_i \in B_i({\bf x}_{-i})$ satisfies $x^*_i = g_i(x^*_i+\bar{x}_{-i},t)$. Note that $g_i(x_T,t)$ is a natural extension of the function $g(x_T)$ defined in \cite{hota2014fragility}. Accordingly, at a fixed tax rate $t$, we have the following result on the monotonicity of the function $g_i(x_T,t)$ with respect to $x_T$. 

\begin{lemma}\label{lemma:gargnyu}
\todo{For a fixed $t > 0$}, $\frac{\partial g_i(x_T,t)}{\partial x_T} < 0$ for $x_T \in \mathcal{I}^t_i$.
\end{lemma}

The proof is analogous to the proof of Lemma 4 in \cite{hota2014fragility}, and thus we omit it. However, $g_i(x_T,t)$ is not always decreasing in $t$ as we will explore later. As a consequence of the above two lemmas, we have the following result.

\vspace{2mm}

\noindent {\bf Proof of Proposition \ref{theorem:cdc_PNEexistence}.} In Lemma \ref{lemma:interval}, we showed that when a player $i$ has a nonzero best response, the total investment in the CPR lies in the interval $\mathcal{I}^t_i$. When $x_T \in S_t$, the rate of return function is monotone, concave and positive. Therefore, the results on the uniqueness and continuity of best responses from Lemma 2 and 3 in \cite{hota2014fragility} carry over to the present setting under taxation. \todo{As a consequence of Brouwer's fixed point theorem \cite{ok2007real}, there exists a fixed point $\mathbf{x}^* \in B(\mathbf{x}^*)$ which corresponds to a PNE.} The uniqueness of PNE follows the monotonicity of nonzero best responses shown in Lemma \ref{lemma:gargnyu}; its proof follows identical arguments as the proof of Theorem 1 in \cite{hota2014fragility}. \hfill \openbox

\subsection{Utilization at a Social Optimum and the PNE}

\noindent {\bf Proof of Proposition \ref{proposition:socopt}.}
Recall from Assumption \ref{assumption:CDCI} that $t < \bar{t}$. Therefore, there exists a player $k$ with $\max_{x_T \in [0,1]} f_k(x_T,t) > 0$. As a result, $\Psi(\mathbf{x}^t_{\mathtt{OPT}},t) > 0$. In the rest of the proof, we omit the superscript $t$ and the second argument from $f$ and $\Psi$ for better readability.

Now, assume on the contrary that $x_{\mathtt{OPT}} > x_{\mathtt{NE}}$. Then there exists a player $i$ with respective CPR investments satisfying $x_{i,\mathtt{OPT}} > x_{i,\mathtt{NE}} > 0$.

First we claim that $f_i(x_{\mathtt{OPT}}) > 0$. Suppose otherwise, and let $j$ be a different player with $f_j(x_{\mathtt{OPT}})>0$.\footnote{Note that such a player always exists; otherwise we have $f_j(x_{\mathtt{OPT}})\leq0$ for every player $j$, which implies $\Psi(\mathbf{x}_{\mathtt{OPT}}) \leq 0$.} Let $\epsilon \in [0,x_{i,\mathtt{OPT}})$, and consider a different strategy profile $\hat{\mathbf{x}}_{\mathtt{OPT}} = (x_{1,\mathtt{OPT}},\ldots,x_{i,\mathtt{OPT}}-\epsilon,\ldots,x_{j,\mathtt{OPT}}+\epsilon,\ldots,x_{n,\mathtt{OPT}})$ with total utilization $x_\mathtt{OPT}$. Then
\begin{align*}
& (x_{i,\mathtt{OPT}})^{\alpha_i} f_i(x_\mathtt{OPT}) + (x_{j,\mathtt{OPT}})^{\alpha_j} f_j(x_\mathtt{OPT}) < (x_{i,\mathtt{OPT}}-\epsilon)^{\alpha_i} f_i(x_\mathtt{OPT}) + (x_{j,\mathtt{OPT}}+\epsilon)^{\alpha_j} f_j(x_\mathtt{OPT})
\\ \implies & \Psi(\mathbf{x}_{\mathtt{OPT}}) < \Psi(\hat{\mathbf{x}}_{\mathtt{OPT}}),
\end{align*}
since $f_i(x_{\mathtt{OPT}}) \leq 0$ and $f_j(x_{\mathtt{OPT}})>0$. This contradicts the optimality of $\mathbf{x}_{\mathtt{OPT}}$. Thus, we must have $f_i(x_{\mathtt{OPT}}) > 0$.

Since $x_{\mathtt{OPT}} > x_{\mathtt{NE}}$ and $f_i(x_{\mathtt{OPT}}) > 0$, $x_{\mathtt{OPT}} \in \mathcal{I}_i$, where $\mathcal{I}_i$ is the interval defined in Lemma \ref{lemma:interval}. From the first order optimality condition for player $i$ at the PNE \eqref{eq:def_foc}, we obtain
\begin{align*}
& x_{i,\mathtt{NE}}f_{i,x}(x_{\mathtt{NE}}) + \alpha_i f_i(x_{\mathtt{NE}}) = 0
\\ \implies & x_{i,\mathtt{OPT}} > x_{i,\mathtt{NE}} = \frac{\alpha_i f_i(x_{\mathtt{NE}})}{-f_{i,x}(x_{\mathtt{NE}})} > \frac{\alpha_if_i(x_{\mathtt{OPT}})}{-f_{i,x}(x_{\mathtt{OPT}})}
\\ \implies & \alpha_if_i(x_{\mathtt{OPT}}) + x_{i,\mathtt{OPT}}f_{i,x}(x_{\mathtt{OPT}}) < 0,
\end{align*}
where $f_{i,x}(x_T) = \frac{\partial f_i}{\partial x_T}(x_T)$, and the second inequality in the second line follows from Lemma \ref{lemma:gargnyu}. 

We now show that for every player $j$ other than $i$, $x_{j,\mathtt{OPT}}^{\alpha_j}f_{j,x}(x_{\mathtt{OPT}}) \leq 0$. For decreasing rate of return functions, this is true since $f_j(\cdot)$ is strictly decreasing in the total investment. On the other hand, for increasing rate of return functions, we have the following two cases.

{\it Case 1: $\max_{x_T\in[0,1]} f_j(x_T) > 0$}. Following the discussion in Lemma \ref{lemma:interval}, we have $x_{\mathtt{NE}} > z_j$ in this case. Therefore, $f_{j,x}(x_{\mathtt{NE}}) < 0$. Furthermore, $f_j(\cdot)$ is concave (following Lemma \ref{lemma:interval}), and $x_{\mathtt{OPT}} > x_{\mathtt{NE}}$, which implies $f_{j,x}(x_{\mathtt{OPT}}) < 0$.

{\it Case 2: $\max_{x_T\in[0,1]} f_j(x_T) \leq 0$}. Following identical arguments as the second paragraph of the proof, we have $x_{j,\mathtt{OPT}} = 0$ in this case.

We are now ready to complete the proof. From the first order optimality condition for the social optimum, we obtain
\begin{align*}
0 = \frac{\partial \Psi}{\partial x_i}\Bigr|_{\mathbf{x} = \mathbf{x}_{\mathtt{OPT}}} & = x_{i,\mathtt{OPT}}^{\alpha_i-1} [x_{i,\mathtt{OPT}}f_{i,x}(x_{\mathtt{OPT}}) + \alpha_i f_i(x_{\mathtt{OPT}})] + \sum^n_{j=1,j\neq i} x_{j,\mathtt{OPT}}^{\alpha_j}f_{j,x}(x_{\mathtt{OPT}}) < 0,
\end{align*}
following the above discussion. This contradicts our initial claim, and we must have $x_{\mathtt{OPT}} \leq x_{\mathtt{NE}}$. \hfill \openbox

\subsection{Support of a PNE and Preliminary Results}

We now define the {\it support} of a PNE.

\begin{definition}\label{def:supp}
The support of the PNE of the game $\Gamma$, denoted $Supp(\Gamma)$, is the set of players who have a nonzero investment in the CPR. In particular, at a tax rate $t$, $Supp(\Gamma) := \{i \in \mathcal{N} | x^t_{\mathtt{NE}} < y^t_i\}$ following Lemma \ref{lemma:interval}. 
\end{definition}

Accordingly, the total investment at the PNE satisfies
\begin{equation}\label{eq:supportPNE}
x^t_{\mathtt{NE}} = \sum_{i \in Supp(\Gamma)} g_i(x^t_{\mathtt{NE}},t).
\end{equation}

\todo{This characterization of PNE investment or utilization will be exploited in many of our subsequent proofs.} \todo{For a player $i \in Supp(\Gamma)$, her investment in the CPR is nonzero. Thus, from Lemma \ref{lemma:interval}, we have $x^t_{\mathtt{NE}} \in \mathcal{I}^t_i$. Recall that for increasing rate of return functions, $\mathcal{I}^t_i = (z^t_i,y^t_i)$, and therefore, $x^t_{\mathtt{NE}} > z^t_i$ for every $i \in Supp(\Gamma)$.} We now present two lemmas that will be useful in several subsequent proofs. 

Let $\Gamma_{1}$ and $\Gamma_{2}$ be two instances of Fragile CPR games with identical resource characteristics and tax rates $t_1$ and $t_2$, respectively. Let the respective total PNE investments be $x^{t_1}_{\mathtt{NE}}$ and $x^{t_2}_{\mathtt{NE}}$. We prove the following result which holds for CPRs with both increasing and decreasing $r(x_T)$.

\begin{lemma}
If $t_1>t_2 \geq 0$, we have $y^{t_1}_i \leq y^{t_2}_i$ for every player $i$ with $\alpha_i \in (0,1]$ and $k_i \in (0,\infty)$. In addition, if $t_1>t_2$ and $x^{t_1}_{\mathtt{NE}} > x^{t_2}_{\mathtt{NE}}$, we have $Supp(\Gamma_1) \subseteq Supp(\Gamma_2)$.
\label{lemma:taxation_supportsize}
\end{lemma}
\begin{proof}
Let $\max_{x \in S_{t_1}} f_i(x,t_1)>0$; otherwise $y^{t_1}_i=0$, and the first statement trivially holds. When $y^{t_1}_i>0$, it follows from Lemma \ref{lemma:interval} that $f_i(y^{t_1}_i,t_1)=0$. When $t_2<t_1$, it is easy to see (from \eqref{eq:def_bt} and \eqref{eq:def_at}) that $S_{t_1} \subseteq S_{t_2}$. Furthermore, note from \eqref{eq:taxedutility} that $f_i$ is decreasing in the second argument $t$ for both increasing and decreasing rate of return functions. Accordingly, $f_i(y^{t_1}_i,t_2)>0$, and therefore, $y^{t_1}_i \leq y^{t_2}_i$.

For the second part of the proof, let $j \in Supp(\Gamma_1)$. From Definition \ref{def:supp}, we have
\begin{align*}
& x^{t_1}_{\mathtt{NE}} < y^{t_1}_j \implies x^{t_2}_{\mathtt{NE}} < x^{t_1}_{\mathtt{NE}} < y^{t_1}_j \leq y^{t_2}_i.
\end{align*}
As a result, $j \in Supp(\Gamma_2)$. This concludes the proof.
\end{proof}

The next lemma shows the monotonicity of $z^t_i$ in $t$ for certain risk preferences.

\begin{lemma}\label{lemma:riskseeking_zt}
Consider a Fragile CPR game with increasing $r(x_T)$, and a player $i$ with $\alpha_i=1$, and let $0 \leq t_2 < t_1 < \bar{t}$. If $k_i < 1$, then $z_i^{t_2} \leq z_i^{t_1}$, and vice versa.
\end{lemma}
\begin{proof}
 When $x_T \in S_t$, the effective rate of return function in \eqref{eq:taxedutility} for player $i$ is given by
\begin{equation}
f_i(x_T,t) \!= \!r(x_T)(1\!-p(x_T)\!)-\!k_ip(x_T)-\!t(k_i-\!1)p(x_T)-\!t.
\end{equation}
Let $z_i^{t_2} > a_{t_1} > 0$; otherwise the result follows directly. According to the first order optimality condition for $z_i^t$, we have $f_{i,x}(z_i^{t_2},t_2) = \frac{\partial f_i}{\partial x_T}(z_i^{t_2},t_2)= 0$. Since $k_i < 1$, and $p(x_T)$ is strictly increasing, it is easy to see that $f_{i,x}(z_i^{t_2},t_1) > 0$ implying $z_i^{t_2} \leq z_i^{t_1}$. The same reasoning applies to the converse.
\end{proof}

Indeed, observe that in Figure \ref{fig:seeking_eff}, $z^t_i$ is increasing in $t$ in accordance with the above lemma. Before we conclude this section, we state Berge's maximum theorem which is used in proving our subsequent results on the continuity of utilization. 

\subsection{Berge's Maximum Theorem}

\begin{theorem}[from \cite{ok2007real}]
Let $\Theta$ and $X$ be two metric spaces, and let $C:\Theta \rightrightarrows X$ be a compact-valued correspondence. Let the function $\Phi: X \times \Theta \to \mathbb{R}$ be jointly continuous in both $X$ and $\Theta$. Define
\begin{align*}
\sigma(\theta) &:= \argmax_{x \in C(\theta)} \Phi(x,\theta),
\text{ and }
\\ \Phi^*(\theta) &:= \max_{x \in C(\theta)} \Phi(x,\theta), \forall \theta \in \Theta.
\end{align*}
If $C$ is continuous at $\theta \in \Theta$, then
\begin{enumerate}
\item $\sigma:\Theta \rightrightarrows X$ is compact-valued, upper hemicontinuous and closed at $\theta$.
\item $\Phi^*:\Theta \rightarrow \mathbb{R}$ is continuous at $\theta$.
\end{enumerate}
\label{theorem:berge}
\end{theorem}

In many instances, the correspondence $C$ takes the form of a parametrized constraint set, i.e., $C(\theta) = \{x \in X| l_j(x,\theta) \leq 0, j \in\{1,2,\ldots,m\}\}$. For this class of constraints, we have the following sufficient conditions for the upper and lower hemicontinuity of $C$ \cite[Theorem 10,12]{hogan1973point}.

\begin{theorem}\label{thm:lhc}
Let $C:\Theta \rightrightarrows X \subseteq \mathbb{R}^{k}$ be given by $C(\theta) = \{x \in X| l_j(x,\theta) \leq 0, j \in \{1,2,\ldots,m\}\}$.
\begin{enumerate}
    \item Let $X$ be closed, and all $l_j$'s be continuous on $X$. Then, $C$ is upper hemicontinunous on $\Theta$.
    \item Let $l_j$'s be continuous and convex in $x$ for each $\theta$. If there exists $(x,\theta)$ such that $l_j(x,\theta) < 0$ for all $j$, then $C$ is lower hemicontinuous at $\theta$, and in some neighborhood of $\theta$.
\end{enumerate}
\end{theorem}

%% file: appendix_network.tex
\section{Proofs Pertaining to CPRs with Network Effects}
\label{appendix:network}

Our goal in this section is to prove Theorem \ref{theorem:cdc_inc_socopt} and Corollary \ref{corollary:cdc_inc_socopt_classical}. We start with some preliminary lemmas that are used in proving the monotonicity and continuity of utilization in the tax rate. The following lemma proves a few useful properties of the function $q_i$ introduced in \eqref{eq:cdc_function_inc_Q}.

\begin{lemma}\label{lemma:cdc_function_q}
The function $q_i$ defined in \eqref{eq:cdc_function_inc_Q} has the following properties.
\begin{enumerate}
\item Let $x_T \in [0,1]$ and $t \in \{t\geq 0|x_T \in (a_t,1]\}$. If $k_i > q_i(x_T,t) > 0$, then $\frac{\partial g_i(x_T,t)}{\partial t} < 0$.
\item Let $q_i(z,t) > 0$ for $z \in (a_t,1]$. Then, $q_i(x_T,t)$ is positive, and is strictly decreasing in $x_T$ for $x_T \in [z,1]$.
\end{enumerate}
\end{lemma}
\begin{proof}
When it is clear from the context, we omit the arguments $x_T$, $t$ and $i$ in the following analysis for better readability. We now state the effective rate of return function under taxation, and compute its derivatives with respect to $x_T$ and $t$. Let $t \in [0,\bar{t})$ and $x_T\in(a_t,1]$. Recall from Assumption \ref{assumption:CDCI} that $r(\cdot)$ is strictly increasing and concave, and $p(\cdot)$ is strictly increasing and convex. From \eqref{eq:taxedutility}, we have
\begin{align}
& f(x_T,t) = (r-t)^\alpha(1-p)-k(1+t)^\alpha p
\\ \implies & f_x(x_T,t) = \frac{\partial f}{\partial x_T}(x_T,t) = \alpha (r-t)^{\alpha-1} r' (1-p) - (r-t)^\alpha p' - k(1+t)^\alpha p'. \label{eq:fx_der}
\end{align}
Differentiating $f(x_T,t)$ with respect to $t$ for $t\in\{t\geq 0|x_T \in (a_t,1]\}$, we obtain
\begin{align}
& f_t(x_T,t) = \frac{\partial f}{\partial t}(x_T,t) = -\alpha (r-t)^{\alpha-1}(1-p) -\alpha k(1+t)^{\alpha-1} p, \qquad \text{and}
\\ & f_{x,t}(x_T,t) = \frac{\partial^2 f}{\partial x_T\partial t}(x_T,t) = -\alpha(\alpha-1)(r-t)^{\alpha-2} \times \nonumber
\\ & \qquad \qquad \qquad r'(1-p) + \alpha (r-t)^{\alpha-1} p' - \alpha k(1+t)^{\alpha-1} p'.
\end{align}
Since $\frac{\partial g}{\partial t} = \frac{ff_{x,t}-f_xf_t}{f^2_x}$, we now compute
\begin{align*}
ff_{x,t} & = -\alpha(\alpha-1)(r-t)^{2\alpha-2}r'(1-p)^2 + \alpha (r-t)^{2\alpha-1}(1-p)p'
\\ & \qquad - \alpha k (r-t)^\alpha (1+t)^{\alpha-1} p'(1-p)- \alpha k (r-t)^{\alpha-1} (1+t)^{\alpha} pp'
\\ & \qquad + \alpha(\alpha-1)k (r-t)^{\alpha-2} r' (1+t)^\alpha p(1-p) + \alpha k^2 (1+t)^{2\alpha-1}pp'.
\end{align*}
Similarly,
\begin{align*}
f_xf_{t} & = -\alpha^2(r-t)^{2\alpha-2}r'(1-p)^2 + \alpha (r-t)^{2\alpha-1}(1-p)p'
\\ & \qquad - \alpha^2k(r-t)^{\alpha-1}r'(1+t)^{\alpha-1}p(1-p) + \alpha k (r-t)^\alpha (1+t)^{\alpha-1} pp'
\\ & \qquad + \alpha k (r-t)^{\alpha-1} (1+t)^{\alpha} p'(1-p) + \alpha k^2 (1+t)^{2\alpha-1}pp'.
\end{align*}
From the above analysis, we obtain
\begin{align}
ff_{x,t} - f_xf_{t} & = \alpha(r-t)^{2\alpha-2}r'(1-p)^2 - \alpha k (r-t)^{\alpha-1} (1+t)^{\alpha} p' - \alpha k (r-t)^{\alpha} (1+t)^{\alpha-1} p' \nonumber
\\ & \qquad + \alpha^2 k (r-t)^{\alpha-1}r'(1+t)^{\alpha-1}p(1-p) + \alpha(\alpha-1)k (r-t)^{\alpha-2} r' (1+t)^\alpha p(1-p) \label{eq:cdc_appendix_firstlemma}
\\ & = \alpha(\alpha-1)k (r-t)^{\alpha-2} r' (1+t)^\alpha p(1-p) - \alpha k (r-t)^{\alpha-1} (1+t)^{\alpha-1} p' (r+1) \nonumber
\\ & \qquad + \alpha (r-t)^{\alpha-1} r' (1-p) \times [(r-t)^{\alpha-1}(1-p) +\alpha k (1+t)^{\alpha-1}p]. \label{eq:cdc_appendix_firstlemma_2}
\end{align}
Since $\alpha < 1$, and $r' > 0$, the first term in \eqref{eq:cdc_appendix_firstlemma_2} is negative. Therefore, a sufficient condition for $ff_{x,t} - f_xf_{t} < 0$ is
\begin{align*}
& k (1+t)^{\alpha-1} p' (r+1)> r' (1-p) [(r-t)^{\alpha-1}(1-p)+\alpha k (1+t)^{\alpha-1}p]
\\ \iff & k (1+t)^{\alpha-1} [(r+1)p' - \alpha r'(1-p)p] > r' (r-t)^{\alpha-1}(1-p)^2
\\ \iff & k > \frac{r'(1-p)^2}{(r+1)p' - \alpha r'(1-p)p} \left(\frac{1+t}{r-t}\right)^{1-\alpha} = q(x_T,t);
\end{align*}
note that $q(x_T,t) > 0$ implies its denominator is positive which is necessary for the above equivalence to hold.

Now, let $l_1 := r'(1-p)^2$ and $l_2 := (r+1)p' - \alpha r'(1-p)p > 0$ be functions of $x_T$ with the argument suppressed for better readability. Then
\begin{align*}
l'_1 & = r''(1-p)^2 - 2r'(1-p)p' < 0, \qquad \text{and}
\\ l'_2 & = (r+1)p'' + r'p' - \alpha r''p(1-p) - \alpha r'p'(1-p) + \alpha r'p'p
\\ & = (r+1)p'' - \alpha r''p(1-p) + r'p'(1-\alpha + 2\alpha p) > 0.
\end{align*}
Suppose $q(z,t) > 0$, i.e., its denominator is positive. Then, for every $x_T \in [z,1]$, the denominator of $q(x_T,t)$ is positive and increasing in $x_T$, and the numerator of $q(x_T,t)$ is decreasing in $x_T$. This concludes the proof.
\end{proof}

Monotonicity of $g_i$ in the tax rate $t$, established above, will be required while proving the monotonicity of the utilization in the tax rate. We now prove several intermediate lemmas towards proving the continuity of utilization in the tax rate.

\subsection{Preliminary results pertaining to continuity of utilization}

We first introduce certain notation, and prove some preliminary lemmas. In appropriate places in this subsection, we treat $y^t_i$, $z^t_i$, and $x^t_{\mathtt{NE}}$ as functions of $t$ (from $[0,\bar{t}) \to [0,1]$), with a slight abuse of notation. Furthermore, we denote the utilization $x_T$ as $x$, and $\frac{\partial f_i}{\partial x}(x,t)$ as $f_{i,x}(x,t)$.

Recall from \eqref{eq:def_at} that $S_t := [a_t,1]$ when $r(x)$ is strictly increasing. Furthermore, $\bar{t}_i := \sup\{t\geq0|\max_{x \in S_t} f_i(x,t)>0\}$. For $t < \bar{t}_i $, $z^t_i := \argmax_{x \in S_t} f_i(x,t)$, and $y^t_i \in (z^t_i,1)$ such that $f_i(y^t_i,t) = 0$. In addition, $f_i(x,t)$ is positive and decreasing for $x \in (z^t_i,y^t_i)$. We now define
\begin{equation}\label{eq:hatzti}
\hat{z}^t_i := \argmax_{x \in [z^t_i,y^t_i]} -[\alpha_if_i(x,t)+f_{i,x}(x,t)]^2.
\end{equation}
Note that at a given $t < \bar{t}_i$, $f_i(x,t)$ is concave, and therefore, $\alpha_if_i(x,t)+f_{i,x}(x,t)$ is strictly decreasing for $x \in [z^t_i,y^t_i]$. Thus, $\hat{z}^t_i = z^t_i$ when $z^t_i = 0$ and $f_{i,x}(0,t) < 0$. Otherwise, $\alpha_if_i(\hat{z}^t_i,t)+f_{i,x}(\hat{z}^t_i,t) = 0$. With the above quantities, we are now ready to define the following function. For a player $i$, $x\in[0,1]$ and $t \in [0,\bar{t}_i)$, let
\begin{equation}\label{eq:cdc_def_ghat_inc}
\hat{g}^N_i(x,t) :=
\begin{cases}
1, \qquad  x \in [0,\hat{z}^t_i),
\\ \frac{\alpha_i f_i(x,t)}{-f_{i,x}(x,t)}, \quad x \in [\hat{z}^t_i,y^t_i),
\\ 0, \qquad \text{otherwise.}
\end{cases}
\end{equation}

\begin{figure}[t!]
        \centering
        \includegraphics[width=6cm,height=4cm]{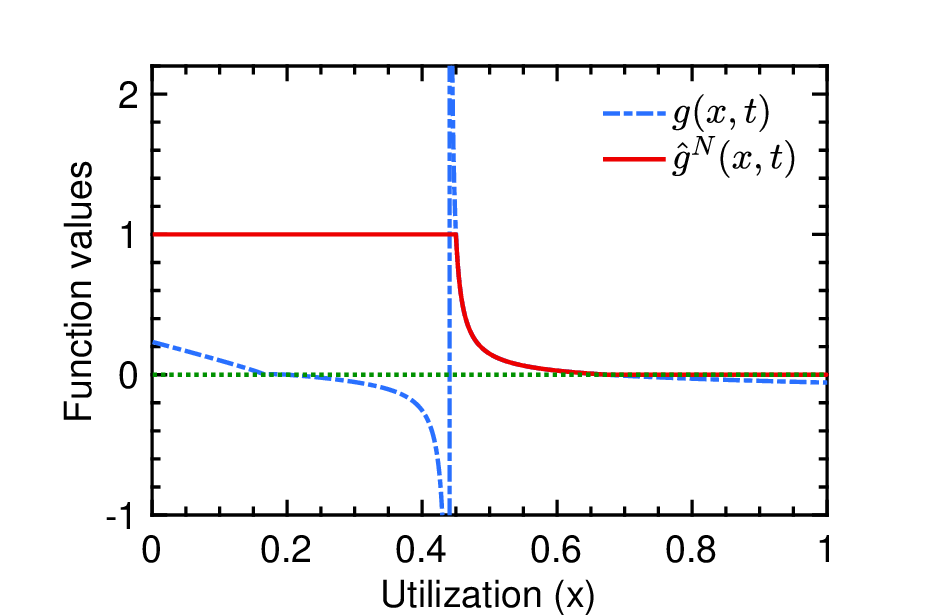}
        \caption{Illustrating the function $\hat{g}^N(x,t)$ for a CPR with $r(x) = 3x+1$, $p(x) = 0.2+0.8x^4$, and $t=1.5$, and a player with $\alpha = 0.3$, $k= 1.5$. In this example, $z^t = 0.4402$ $\hat{z}^t = 0.4502$, and $y^t=0.6737$. Note that $g(x,t)$ is undefined at $x=z^t$.}
        \label{fig:hatz}
\end{figure}

Note that $\hat{g}^N_i(x,t)$ is well-defined. It follows from \eqref{eq:hatzti} that when $\hat{z}^t_i > 0$, the maximum value of $\frac{\alpha_i f_i(x,t)}{-f_{i,x}(x,t)} = 1$ for $x\in[\hat{z}^t_i,y^t_i]$ occurs at $x = \hat{z}^t_i$. As a result, $\hat{g}^N_i(x,t)$ is bounded.\footnote{If we had defined the range of $x$ to be $[z^t_i,y^t_i)$ in the second line of \eqref{eq:cdc_def_ghat_inc}, then the denominator of $\hat{g}^N_i(x,t)$ would be $0$ at $x=z^t_i$.} In Figure \ref{fig:hatz}, we illustrate the shape of the function $\hat{g}^N_i(x,t)$, and how it compares with $g_i(x,t)$ defined in \eqref{eq:defg} for the CPR with the same characteristics as Example \ref{ex:cont_multiple}. Note that the denominator of $g_i(x,t)$ is $0$ at $x=z^t_i$, while $\hat{g}^N_i(x,t)$ is bounded for $x\in[0,1]$ as it is defined in terms of $\hat{z}^t_i$.

We first establish the continuity of $z^t_i$,$\hat{z}^t_i$, and $y^t_i$, and then prove the (joint) continuity of $\hat{g}^N_i(\cdot,\cdot)$.

\begin{lemma}\label{lemma:cont_hatz}
When viewed as functions of $t$, $z^t_i$,$\hat{z}^t_i$, and $y^t_i$ are continuous in $t$ for $t \in [0,\bar{t}_i)$.
\end{lemma}
\begin{proof}
With a slight abuse of notation, we view the set $S_t$ as a compact-valued correspondence $S: [0,\bar{t}_i) \rightrightarrows [0,1]$ with $S(t) := \{x \in [0,1]| t - r(x) \leq 0\}$. Since $r(x)$ is continuous, concave, and for every $t < \bar{t}_i$, $t - r(1) < 0$, it follows from Theorem \ref{thm:lhc} that $S(t)$ is both upper and lower hemicontinuous. Note that $f_i(x,t)$ is defined for $x \in [0,1]$, $t \in [0,\bar{t}_i)$, and is jointly continuous in $x$ and $t$. Recall further that $z^t_i := \argmax_{x \in S(t)} f_i(x,t)$; $z^t_i$ is single-valued since $f_i(x,t)$ is strictly concave in $x \in S(t)$ at a given $t$. Therefore, following Berge's maximum theorem, $z^t_i$ is continuous in $t$.

Recall that $y^t_i \in [z^t_i,1]$ such that $f_i(z^t_i,t) = 0$. Furthermore, $f_i(x,t)$ is strictly decreasing in $x$ for $x \in (z^t_i,1]$. Therefore, we can alternatively let $y^t_i := \argmax_{x \in [z^t_i,1]} -(f_i(x,t))^2$. Since $z^t_i$ is continuous in $t$, the correspondence $t \rightrightarrows [z^t_i,1]$ is continuous following Theorem \ref{thm:lhc}. Berge's maximum theorem now implies that $y^t_i$ is continuous in $t$.

Following the above discussion, we have $t \rightrightarrows [z^t_i,y^t_i]$ as continuous. From its definition \eqref{eq:hatzti}, $\hat{z}^t_i$ is the unique maximizer of a function that is jointly continuous in both $x$ and $t$. Once again, from Berge's maximum theorem, we conclude that $\hat{z}^t_i$ is continuous in $t$.
\end{proof}

\begin{lemma}\label{lemma:cdc_inc_cont}
The function $\hat{g}_i(x,t), x \in [0,1], t \in [0,\bar{t}_i)$ defined in \eqref{eq:cdc_def_ghat_inc} is jointly continuous in $x$ and $t$.
\end{lemma}
\begin{proof}
First observe that at a given $t$, $\hat{g}_i(x,t)$ is continuous and monotone in $x$ for $x \in [0,1]$. In particular, $\hat{g}_i(y^t_i,t) = f_i(y^t_i,t)=0$, and the monotonicity follows from Lemma \ref{lemma:gargnyu}. Following \cite{kruse1969joint}, it now suffices to show that $\hat{g}_i(x,t)$ is continuous in $t$ at a given $x$. However, this is true because $\hat{z}^t_i$ and $y^t_i$ are continuous in $t$ following Lemma \ref{lemma:cont_hatz}, together with $\frac{\alpha_i f_i(\hat{z}^t_i,t)}{-f_{i,x}(\hat{z}^t_i,t)} = 1$, and $\frac{\alpha_i f_i(y^t_i,t)}{-f_{i,x}(y^t_i,t)} = 0$.
\end{proof}

We now show that the total PNE investment can be stated as a maximizer of a function that is continuous in both the total investment and the tax rate.

\begin{lemma}\label{lemma:cdc_inc_cont2}
For $x \in [0,1], t \in [0,\min_{i \in \mathcal{N}} \bar{t}_i)$, define $$h^N(x,t):=-[x-\sum_{i\in\mathcal{N}}\hat{g}^N_i(x,t)]^2.$$ Then, at a given $t$, $\argmax_{x \in [0,1]} h^N(x,t)$ is single-valued, and is equal to $x^t_{\mathtt{NE}}$.
\end{lemma}

\begin{proof}
From its definition, $h^N(x,t) \leq 0$. Suppose there exists $x^t \in [0,1)$ such that $h^N(x^t,t) = 0$, or equivalently $x^t = \sum_{i\in\mathcal{N}} \hat{g}^N_i(x^t,t)$. First we claim that $x^t > \hat{z}^t_j$ for every player $j$. If this is not the case, then for a player $j$ with $x^t \leq \hat{z}^t_j$, we have $\hat{g}^N_j(x^t,t) = 1$ which implies $x^t < \sum_{i\in\mathcal{N}} \hat{g}^N_i(x^t,t)$.

Now consider the strategy profile $\{x^t_j\}_{j \in \mathcal{N}}$ where $x^t_j = \hat{g}^N_j(x^t,t)$ for each player $j$. Consider a player $j$ with $x^t \geq y^t_j$. Then, $x^t_j = \hat{g}^N_j(x^t,t) = 0$. Following Lemma \ref{lemma:interval}, the strategy of player $j$, $x^t_j$, is her best response. Now suppose $x^t < y^t_j$. From the definition of $\hat{g}^N$, we have $x^t_j f_{j,x} (x^t,t) + \alpha_j f_{j} (x^t,t) = 0$. Following \eqref{eq:def_foc}, the investment of player $j$ satisfies the first order optimality condition for her utility. Furthermore, the proof of Lemma 2 in \cite{hota2014fragility} showed that the utility of player $j$ is strictly concave in the range of investments which contains the investment at which the first order optimality condition is satisfied. Therefore, $x^t_j$ is the unique best response of player $j$ for the given strategies of others. Thus, $\{x^t_i\}_{i \in \mathcal{N}}$ corresponds to a PNE strategy profile. Conversely, it is easy to see that at any PNE strategy profile, $h^N(x^t_{\mathtt{NE}},t)=0$.

Recall that a PNE exists, and is unique. Following Theorem 1 in \cite{hota2014fragility}, the total investment at the PNE is unique as well. Therefore, there is a unique $x = x^t_{\mathtt{NE}}$ with $h^N(x,t)=0$, which also maximizes $h^N(x,t)$ at a given $t$.
\end{proof}

With the above preliminary results in place, we are now ready to prove Theorem \ref{theorem:cdc_inc_socopt}.

\subsection{Proof of Theorem \ref{theorem:cdc_inc_socopt}}

\begin{proof}
\noindent {\bf Part 1 (monotonicity).}
Assume on the contrary that $x^{t_1}_{\mathtt{NE}} > x^{t_2}_{\mathtt{NE}} > 0$. According to Lemma~\ref{lemma:taxation_supportsize}, we have $Supp(\Gamma_1) \subseteq Supp(\Gamma_2)$. From the characterization of PNE in equation \eqref{eq:supportPNE}, we obtain
\begin{align}
& x^{t_1}_{\mathtt{NE}} > x^{t_2}_{\mathtt{NE}} \nonumber
\\ \implies & \sum_{j \in Supp(\Gamma_1)} g_j(x^{t_1}_{\mathtt{NE}},t_1) > \sum_{j \in Supp(\Gamma_2)} g_j(x^{t_2}_{\mathtt{NE}},t_2) \nonumber
\\ \implies & \sum_{j \in Supp(\Gamma_1)} g_j(x^{t_1}_{\mathtt{NE}},t_1) > \sum_{j \in Supp(\Gamma_1)} g_j(x^{t_2}_{\mathtt{NE}},t_2). \label{eq:taxmonotonecontradiction1}
\end{align}
In the remainder of the proof, our goal is to contradict the inequality in equation~\eqref{eq:taxmonotonecontradiction1}. In particular, for each player $j \in Supp(\Gamma_1)$, we show that $g_j(x^{t_1}_{\mathtt{NE}},t_1) < g_j(x^{t_2}_{\mathtt{NE}},t_2)$.

Consider a player $j \in Supp(\Gamma_1)$. From Lemma \ref{lemma:interval}, $x^{t_1}_{\mathtt{NE}}<y^{t_1}_j$. Furthermore, from Lemma \ref{lemma:taxation_supportsize}, we have $y^{t_1}_j \leq y^{t_2}_j$. Together with our assumption, we obtain
$$ x^{t_2}_{\mathtt{NE}}<x^{t_1}_{\mathtt{NE}}<y^{t_1}_j\leq y^{t_2}_j \implies [x^{t_2}_{\mathtt{NE}},x^{t_1}_{\mathtt{NE}}]\subset\mathcal{I}^{t_2}_{j}.$$
From the monotonicity of $g_j(x_T,t)$ in $x_T$ in Lemma~\ref{lemma:gargnyu}, we obtain $g_j(x^{t_1}_{\mathtt{NE}},t_2) < g_j(x^{t_2}_{\mathtt{NE}},t_2)$.

It is now sufficient to show that $g_j(x^{t_1}_{\mathtt{NE}},t_1) < g_j(x^{t_1}_{\mathtt{NE}},t_2)$. Since $x^{t_1}_{\mathtt{NE}} > x^{t_2}_{\mathtt{NE}} > a_{t_1}$, part two of Lemma \ref{lemma:cdc_function_q} yields
$$ q_j(x^{t_2}_{\mathtt{NE}},t_1) > q_j(x^{t_1}_{\mathtt{NE}},t_1). $$
Furthermore, from its definition, $q_j(x,t)$ is strictly increasing in $t$. Thus, for $t \in [t_2,t_1]$,
$$ q_j(x^{t_1}_{\mathtt{NE}},t_1) \geq q_j(x^{t_1}_{\mathtt{NE}},t);$$
note that $x^{t_1}_{\mathtt{NE}} > a_{t_1} \geq a_t$ for $t \in [t_2,t_1]$. Combining these observations, we obtain
\begin{align*}
& k_j > q_j(x^{t_2}_{\mathtt{NE}},t_1) > q_j(x^{t_1}_{\mathtt{NE}},t_1) \geq q_j(x^{t_1}_{\mathtt{NE}},t) > 0
\\ \implies & \frac{\partial g_j(x^{t_1}_{\mathtt{NE}},t)}{\partial t} < 0,
\end{align*}
for $t \in [t_2,t_1]$ (following the first part of Lemma \ref{lemma:cdc_function_q}). Therefore, $g_j(x^{t_1}_{\mathtt{NE}},t_1) < g_j(x^{t_1}_{\mathtt{NE}},t_2)$, which contradicts \eqref{eq:taxmonotonecontradiction1}.

\noindent {\bf Part 2 (continuity).} Let $C: [0,\min_{i \in \mathcal{N}} \bar{t}_i) \rightrightarrows [0,1]$ such that $C(t) = [0,1]$ for $t \in [0,\min_{i \in \mathcal{N}} \bar{t}_i)$. From its definition, $C$ is compact-valued, and is both upper and lower hemicontinuous at every $t \in [0,\min_{i \in \mathcal{N}} \bar{t}_i)$. From Lemma \ref{lemma:cdc_inc_cont}, $h^N(x,t)$ is jointly continuous in $x$ and $t$. Following Berge's maximum theorem, $\argmax_{x\in C(t)} h^N(x,t)$ is upper hemicontinuous. From Lemma \ref{lemma:cdc_inc_cont2}, $h^N(x,t)$ is single-valued. Therefore, $x_{\mathtt{NE}}(t)$ is continuous in $t$ for $t \in [0,\min_{i \in \mathcal{N}} \bar{t}_i)$.

\noindent {\bf Part 3 (existence of suitable tax rates).} The third statement now follows from the extreme value theorem. For the fourth statement, note from the definition of $\bar{t}_i$ that it is the smallest tax rate at which the maximum value of $f_i(x_T,\bar{t}_i) = 0, x_T \in S_{\bar{t}_i}$. From the definitions of $z^t_i$ and $y^t_i$, we have $\lim_{t \uparrow \bar{t}_i} z^t_i = \lim_{t \uparrow \bar{t}_i} y^t_i$. The continuity of $x_{\mathtt{NE}}(t)$, and the intermediate value theorem now suffice. The fifth statement now follows from Proposition \ref{proposition:socopt} which states that $x^0_{\mathtt{OPT}} \leq x^0_{\mathtt{NE}}$.
\end{proof}

\subsection{Proof of Corollary \ref{corollary:cdc_inc_socopt_classical}}
\begin{proof}
Assume on the contrary that $x^{t_1}_{\mathtt{NE}} > x^{t_2}_{\mathtt{NE}}$. Following analogous arguments as the first part of the proof of Theorem \ref{theorem:cdc_inc_socopt}, it suffices to show that for a player $j \in Supp(\Gamma_1)$, $g_j(x^{t_1}_{\mathtt{NE}},t_1) < g_j(x^{t_1}_{\mathtt{NE}},t_2)$.

Note that $x^{t_1}_{\mathtt{NE}} > a_{t_1} \geq a_{t_2}$. Therefore, $f_j(x^{t_1}_{\mathtt{NE}},t)$ is defined for $t \in [t_2,t_1]$. Since $\alpha_j=1$, we have
\begin{align*}
f_j(x_T,t) & = (r(x_T)-t)(1-p(x_T))-k_j(1+t)p(x_T) - t,
\\ f_{j,x}(x_T,t) & = \frac{\partial f_j}{\partial x_T}(x_T,t) = r'(x_T)(1-p(x_T)) -r(x_T)p'(x_T)-k_jp'(x_T)-tp'(x_T)(k_j-1).
\end{align*}
It is easy to see that $\frac{\partial f_j}{\partial t}(x^{t_1}_{\mathtt{NE}},t) < 0$. Furthermore, when $k_j \geq 1$, $\frac{\partial^2 f_j}{\partial x_T \partial t}(x^{t_1}_{\mathtt{NE}},t)\leq0$ for $t \in [t_2,t_1]$. Therefore, $g_j(x^{t_1}_{\mathtt{NE}},t)$ is decreasing in $t$ for $t \in [t_2,t_1]$. The monotonicity part now follows from identical arguments as the proof of Theorem \ref{theorem:cdc_inc_socopt}.

Recall now that $z^t_i := \argmax_{x \in [a_t,1]} f_i(x,t)$, and Lemma \ref{lemma:riskseeking_zt} states that as the tax rate increases, $z^t$ decreases for every player. \todo{In addition, since all players have identical $\alpha$ and $k$, it follows from \cite[Proposition 4]{hota2014fragility} that $x^0_{\mathtt{OPT}}$ is equal to the investment by a single player when she invests in isolation. In that case, $x^0_{\mathtt{OPT}}$ satisfies the first order optimality condition 
$$ \alpha f(x^0_{\mathtt{OPT}}) + x^0_{\mathtt{OPT}} f_x (x^0_{\mathtt{OPT}}) = 0.$$
Consequently, following the proof of Lemma 1, in particular equation (11), we have $x^0_{\mathtt{OPT}} \in \mathcal{I}^0$, or $z^0 <x^0_{\mathtt{OPT}}$.} Therefore, $\bar{x} = \lim_{t \uparrow \bar{t}} z^t < z^0 < x^0_{\mathtt{OPT}}$. The result now follows from Theorem \ref{theorem:cdc_inc_socopt}.
\end{proof}

%% file: appendix_congestion.tex
\section{Proofs Pertaining to CPRs with Congestion Effects}
\label{appendix:congestion}

Our approach for proving Theorem \ref{theorem:main_dec} is along similar lines as Appendix \ref{appendix:network}. We start with a lemma which holds for the general form of the utility function \eqref{eq:taxedutility} with $\alpha_i \in (0,1]$.

\begin{lemma}\label{lemma:decreasingDGT}
Let $r(x_T)$ be decreasing in $x_T$. For a player $j$ and a given $x_T \in [0,1]$, let $T^{x_T}_j := \{t\in[0,\bar{t})|f_j(x_T,t) > 0\}$. Let $g_j(x_T,t)$ be the function defined in \eqref{eq:defg}. Then, $\frac{\partial g_j(x_T,t)}{\partial t} < 0$ for $t\in T^{x_T}_j$.
\end{lemma}
\begin{proof}
Let $\bar{k} := k(1+t)^\alpha$, and $\bar{k}_t := k\alpha(1+t)^{\alpha-1}$. From \eqref{eq:cdc_appendix_firstlemma}, we obtain
\begin{align*}
ff_{x,t} - f_tf_x & = \alpha (r-t)^{2\alpha-2} r' (1-p)^2 - (r-t)^\alpha \bar{k}_t p'
\\ & \quad - \alpha(1-\alpha) (r-t)^{\alpha-2} r' \bar{k} p(1-p) + \alpha (r-t)^{\alpha-1} r' \bar{k}_t (1-p) p - \alpha (r-t)^{\alpha-1} \bar{k} p'
\\ & = \alpha (1-p) r' (r-t)^{\alpha-2} [(r-t)^\alpha (1-p) - (1-\alpha) \bar{k}p
\\ & \quad + (r-t)\bar{k}_t p] - \alpha (r-t)^{\alpha-1} \bar{k} p' - (r-t)^\alpha \bar{k}_t p'
\\ & = \alpha (1-p) r' (r-t)^{\alpha-2} [f + \alpha \bar{k}p + (r-t)\bar{k}_t p]
\\ & \quad - \alpha (r-t)^{\alpha-1} \bar{k} p' - (r-t)^\alpha \bar{k}_t p' < 0,
\end{align*}
when $r' < 0$ and $f > 0$. This concludes the proof.
\end{proof}

In other words, at a given utilization level $x_T$, the function $g_j(x_T,t)$ is strictly decreasing in the tax rate $t$ over the range of tax rates at which the effective rate of return remains positive. This property will be key towards proving the monotonicity of utilization in the tax rate.

In order to prove the continuity of utilization, we rely on Berge's maximum theorem as before. In order to apply Berge's maximum theorem, we need to express the total PNE investment $x^t_{\mathtt{NE}}$ as the unique maximizer of a function that is jointly continuous in the total investment and the tax rate. 

First we define the following function. For a player $i$, $x_T\in[0,1]$ and $t \in [0,\bar{t})$, let
\begin{equation}\label{eq:cdc_def_ghat_dec}
\hat{g}_i(x_T,t) :=
\begin{cases}
\frac{\alpha_i f_i(x_T,t)}{-f_{i,x}(x_T,t)}, \qquad x_T \in [0,y^t_i)
\\ 0, \qquad \text{otherwise.}
\end{cases}
\end{equation}
Note that $\hat{g}_i(x_T,t)$ is bounded\footnote{For $x_T \in [0,1]$, and nonempty $T^{x_T}_j := \{t\in[0,\bar{t})|f_j(x_T,t) > 0\}$, Lemma \ref{lemma:decreasingDGT} shows that $\frac{\partial g_j(x_T,t)}{\partial t} < 0$ for $t\in T^{x_T}_j$. In this case, $\frac{\alpha_i f_i(x_T,t)}{-f_{i,x}(x_T,t)} < \frac{\alpha_i f_i(x_T,0)}{-f_{i,x}(x_T,0)}$. Furthermore, $f_{i,x}(x_T,0)$ (with expression \eqref{eq:fx_der}) is strictly smaller than $0$ under Assumption \ref{assumption:CDCI}.}, and therefore well-defined. In the following lemma, we prove the (joint) continuity of $\hat{g}_i(\cdot,\cdot)$.

\begin{lemma}\label{lemma:cdc_dec_cont}
The function $\hat{g}_i(x_T,t), x_T \in [0,1], t \in [0,\bar{t})$ defined in \eqref{eq:cdc_def_ghat_dec} is jointly continuous in $x_T$ and $t$.
\end{lemma}
\begin{proof}
First observe that at a given $t$, $\hat{g}_i(x_T,t)$ is continuous and monotone in $x_T$ for $x_T \in [0,1]$; since $f_i(y^t_i,t)=0$, we have $\hat{g}_i(y^t_i,t) = 0$, and the monotonicity follows from Lemma \ref{lemma:gargnyu}. Following \cite{kruse1969joint}, it now suffices to show that $\hat{g}_i(x_T,t)$ is continuous in $t$ at a given $x_T$.

Since $f_i(x_T,t)$ is strictly decreasing in $t$, the condition $x_T \in [0,y^t_i)$ is equivalent to $t \in [0,\hat{t}^{x_T}_i)$, where $\hat{t}^{x_T}_i := \min\{t \in [0,\bar{t}]: f_i(x_T,t) \leq 0\}$. For $t \in [0,\hat{t}^{x_T}_i)$, $\hat{g}_i(x_T,t)$ is continuous in $t$ as both numerator and denominator are continuous in $t$. For $t \geq \hat{t}^{x_T}_i$, $\hat{g}_i(x_T,t) = 0$. Furthermore, when $\hat{t}^{x_T}_i > 0$, $f_i(x_T,\hat{t}^{x_T}_i) = 0$. Thus, $\hat{g}_i(x_T,t)$ is continuous in $t$ at a given $x_T \in [0,1]$.
\end{proof}

We now show that the total PNE investment can be stated as a maximizer of a function that is continuous in both the total investment and the tax rate.

\begin{lemma}\label{lemma:cdc_dec_cont2}
Define
$$h^C(x_T,t):=-[x_T-\sum_{i\in\mathcal{N}}\hat{g}_i(x_T,t)]^2, x_T \in [0,1], t \in [0,\bar{t}).$$
Then, at a given $t$, $\argmax_{x_T \in [0,1]} h^C(x_T,t)$ is single-valued, and is equal to $x^t_{\mathtt{NE}}$.
\end{lemma}

The proof follows from identical arguments as the proof of Lemma \ref{lemma:cdc_inc_cont2} in Appendix \ref{appendix:network}, and is omitted. 

\vspace{2mm}

\noindent {\bf Proof of Theorem \ref{theorem:main_dec}.}
The proof of monotonicity relies on similar arguments as the proof of Theorem \ref{theorem:cdc_inc_socopt}. Specifically, a contradiction to \eqref{eq:taxmonotonecontradiction1} is obtained from Lemma \ref{lemma:gargnyu} and Lemma \ref{lemma:decreasingDGT} which for every player $j \in Supp(\Gamma_1)$ imply $g_j(x^{t_1}_{\mathtt{NE}},t_1) < g_j(x^{t_2}_{\mathtt{NE}},t_1)$ and $g_j(x^{t_2}_{\mathtt{NE}},t_1) < g_j(x^{t_2}_{\mathtt{NE}},t_2)$, respectively. We omit the details in the interest of space; the complete proof can be found in \cite{hota2016controlling}.

Now we focus on the proof of continuity. Consider a set-valued map or correspondence $C: [0,\bar{t}) \rightrightarrows [0,1]$ such that $C(t) = [0,1]$ for every $t \in [0,\bar{t})$. From its definition, $C$ is compact-valued, and is both upper and lower hemicontinuous at every $t \in [0,\bar{t})$. From Lemma \ref{lemma:cdc_dec_cont}, $h^C(x_T,t)$ is jointly continuous in $x_T$ and $t$. Following Berge's maximum theorem (see Appendix \ref{appendix:PNE}), the set-valued map $\argmax_{x_T\in C(t)} h^C(x_T,t)$ is upper hemicontinuous. From Lemma \ref{lemma:cdc_dec_cont2}, we have $\argmax_{x_T\in C(t)} h^C(x_T,t) = \{x_{\mathtt{NE}}(t)\}$, i.e., the set-valued map is in fact single-valued. Therefore, $x_{\mathtt{NE}}(t)$ is continuous in $t$ for $t \in [0,\bar{t})$.

We now show that $x_{\mathtt{NE}}(t)$ is continuous at $t=\bar{t}$. From the strict monotonicity and continuity of $f(\cdot,\cdot)$ and the definition of $\bar{t}$, we have $\max_{i \in \mathcal{N}} f_i(0,\bar{t})=0$. Therefore, $x_{\mathtt{NE}}(\bar{t}) = 0$. Now, recall from Lemma \ref{lemma:interval} and Definition \ref{def:supp} that $x_{\mathtt{NE}}(t) < \max_{i \in \mathcal{N}} y^t_i$ at any tax rate $t$. Furthermore, as $t \uparrow \bar{t}$, we have
$$ \max_{i \in \mathcal{N}} f_i(0,t) \to 0 \implies \max_{i \in \mathcal{N}} y^t_i \to 0 \implies x_{\mathtt{NE}}(t) \to 0.$$

The third statement follows from the continuity property and the extreme value theorem. Furthermore, Proposition \ref{proposition:socopt} states that $x^0_{\mathtt{OPT}} \leq x^0_{\mathtt{NE}}$. Therefore, the fourth part follows from the monotonicity and continuity of utilization in the tax rate shown above. \hfill \openbox

%% file: appendix_diff.tex
\section{Proofs Pertaining to Differentiated Tax Rates}
\label{appendix:diff}

\noindent {\bf Proof of Proposition \ref{prop:sensitivity_uniform}.} (Sketch) First we observe that the effective rate of return $f_i(x,t)$ defined in \eqref{eq:taxsensitivity} for the case with player-specific $\gamma_i$'s is analogous to the effective rate of return in a game without taxes with a modified rate of return $r(x_T) - \gamma_i t$ and a modified index of loss aversion $k(1+\gamma_i t)$. Consequently, if $r(x_T)$ and $p(x_T)$ satisfy Assumption \ref{assumption:CDCI}, then for $t \in [0,\bar{t})$, so do $r(x_T) - \gamma_i t$ and $p(x_T)$. Accordingly, Lemma \ref{lemma:interval} and Lemma \ref{lemma:gargnyu} continue to hold for each player $i$. Furthermore, since the proof of Proposition \ref{proposition:socopt} relies on Lemma \ref{lemma:interval} and Lemma \ref{lemma:gargnyu}, we have $x_{\mathtt{OPT}}(t) \leq x_{\mathtt{NE}}(t)$ at a given tax rate $t \in [0,\bar{t})$ in the case with heterogeneous $\gamma_i$'s.

For CPRs with increasing rate of return functions, following an analogous approach to the proof of Corollary \ref{corollary:cdc_inc_socopt_classical}, we can show that the function $g_j(x_T, t)$ is a decreasing function of $t$ over a suitable domain; this result is a consequence of our assumption that $\alpha=1$ and $k>1$. As a consequence, utilization is monotonically decreasing in the tax rate. Furthermore, Lemmas \ref{lemma:cont_hatz}, \ref{lemma:cdc_inc_cont} and \ref{lemma:cdc_inc_cont2} continue to hold with analogous arguments which imply that utilization is continuous as a function of $t$ for $t \in [0,\min_{i \in \mathcal{N}} \bar{t}_i)$ where $\bar{t}_i = \sup\{t\geq0|\max_{x \in [0,1]} f_i(x,t) > 0\}$. Therefore, if $x_\mathtt{OPT}(0) > x_\mathtt{NE}(t^*)$, then there exists a tax rate such that utilization at the NE is equal to $x_\mathtt{OPT}(0)$. Now for CPRs with decreasing rate of return functions, it can be shown that Lemmas \ref{lemma:decreasingDGT} and \ref{lemma:cdc_dec_cont} continue to hold in the case with player-specific $\gamma_i$'s. As a result, Theorem \ref{theorem:main_dec} holds, and there exists a uniform tax rate such that any desired utilization in the range $[0,x_{\mathtt{NE}}(0)]$ can be achieved. \hfill \openbox

\noindent {\bf Proof of Proposition \ref{prop:diff}.} Let $\Gamma_H \in \mathbf{\Gamma}_m$ be a Fragile CPR game where the effective tax rates are heterogeneous. Without loss of generality, let $0 \leq t_1 \leq t_2 \leq \ldots \leq t_n$, with $\sum^n_{i=1} t_i = n t_m$. Furthermore, let the utilizations at the respective PNEs of $\Gamma_H$ and $\Gamma_M$ be $x_H$ and $x_M$.

Suppose $x_H = 0$. Then, we have $\hat{f}(x_T) - t_1 v(x_T) \leq 0$ for $x_T \in S_{t_1}$. Since $t_m \geq t_1$, we also have  $\hat{f}(x_T) - t_m v(x_T) \leq 0$ for $x_T \in S_{t_m}$, which implies $x_M = 0$. Since $t_m < \bar{t}_m$, we must have $x_M > 0$, and thus, the case $x_H = 0$ does not arise.

Therefore, $x_H > 0$. For $j \notin Supp(\Gamma_H)$, we have
\begin{align*}
\hat{f}(x_H) - t_j v(x_H) & \leq 0 \implies \hat{f}(x_H) - t v(x_H) \leq 0,
\end{align*}
for every $t \geq t_j$. Therefore, $Supp(\Gamma_H)$ consists of a set of players with smallest effective tax rates. Since $x_H > 0$, player $1 \in Supp(\Gamma_H)$. From equation \eqref{eq:supportPNE} for $\Gamma_H$, we have
\begin{align*}
x_H & = \sum_{i \in Supp(\Gamma_H)} g_i(x_H,t_i) = \sum^n_{i=1} \max(g_i(x_H,t_i),0)
\\ & = \sum^n_{i=1} \max\left(\frac{\hat{f}(x_H) - t_i v(x_H)}{-\hat{f}'(x_H) + t_i v'(x_H)},0 \right)
\\ & =: \sum^n_{i=1} \max(h_{x_H}(t_i),0),
\end{align*}
where $h_{x_H}(\cdot)$ is a function of $t$ at a given total investment $x_H$. Note that, since the players are loss averse, we have $v'(x_H)  = (k-1)p'(x_H) > 0$. As a result, for $t \geq t_1$, the numerator of $h_{x_H}(t)$ is strictly decreasing in $t$, while the denominator is strictly increasing in $t$. 

We now define an interval $\mathcal{J} \subseteq [t_1,n t_m]$ as follows. If $h_{x_H}(n t_m) > 0$, then $\mathcal{J} = [t_1,n t_m]$. Otherwise, $\mathcal{J} = [t_1,t^u)$, where $t^u \leq n t_m$ is the unique effective tax rate at which $h_{x_H}(t^u) = 0$, and every player $i \in Supp(\Gamma_H)$ satisfies $t_i \in \mathcal{J}$. For $t \in \mathcal{J}$, we have $\hat{f}(x_H) - t v(x_H) > 0$ and $-\hat{f}'(x_H) + t v'(x_H) > 0$, which implies
\begin{align}
\hat{f}(x_H) v'(x_H) & > t v(x_H) v'(x_H) > \hat{f}'(x_H) v(x_H). \label{eq:convex_h}
\end{align}
For $t \in \mathcal{J}$, straightforward calculations yield
\begin{align*}
h'_{x_H}(t) & = \frac{(\hat{f}'(x_H)v(x_H) - \hat{f}(x_H)v'(x_H))}{(-\hat{f}'(x_H)+t v'(x_H))^2} < 0,
\\ h''_{x_H}(t) & = \frac{-2v'(x_H)(\hat{f}'(x_H)v(x_H) - \hat{f}(x_H)v'(x_H))}{(-\hat{f}'(x_H)+ t v'(x_H))^3}.
\end{align*}
Following \eqref{eq:convex_h}, we have $h''_{x_H}(t) > 0$ for $t \in \mathcal{J}$. Therefore, $\max(h_{x_H}(t,0))$ is continuous and convex for $t \in [t_1,n t_m]$. Applying Jensen's inequality, we obtain
\begin{align*}
x_H = \sum^n_{i=1} \max(h_{x_H}(t_i),0) \geq n \max(h_{x_H}(t_m),0).
\end{align*}

We now consider two cases. First, suppose $h_{x_H}(t_m) \leq 0$. Note that $-\hat{f}'(x_H) + t_m  v'(x_H) > 0$ (since $t_m \geq t_1$ and $v'(x_H) > 0$). Thus, we have $\hat{f}(x_H) - t_m  v(x_H) \leq 0$. When $r(x_T)$ is decreasing, it is easy to see that $\hat{f}(x_T) - t_m v(x_T) < 0$ for $x_T \in (x_H,1]$. For an increasing and concave $r(x_T)$, $\hat{f}(x_T) - t_m v(x_T)$ is strictly concave in $x_T$. Since $\hat{f}'(x_H) - t_m v'(x_H) < 0$ and $\hat{f}(x_H) - t_m v(x_H) \leq 0$, we have $\hat{f}(x_T) - t_m v(x_T) < 0$ for $x_T \in (x_H,1]$. Thus, $x_M \leq x_H$.

Now suppose $h_{x_H}(t_m) > 0$, i.e., $\hat{f}(x_H) - t_m v(x_H) > 0$ and $\hat{f}'(x_H) - t_m v'(x_H) < 0$. Assume on the contrary that $x_M > x_H$. Thus, we have $[x_H,x_M] \subset \mathcal{I}_m$, where $\mathcal{I}_m$ is the interval defined in Lemma~\ref{lemma:interval} for a player $m$ with effective tax rate $t_m$. Following Lemma~\ref{lemma:gargnyu}, we obtain
\begin{align*}
x_H \geq n h_{x_H}(t_m) = n g_m(x_H,t_m) > n g_m(x_M,t_m) = x_M,
\end{align*}
which is a contradiction. Therefore, $x_H \geq x_M$. \hfill \openbox